\documentclass[10pt,aps,prapplied,twocolumn,longbibliography,nobibnotes,showkeys]{revtex4-2}

\usepackage{graphicx}
\usepackage{dcolumn}
\usepackage{bm}
\usepackage[version=3]{mhchem}
\usepackage{color}
\usepackage{booktabs}

\begin{document}
\title{Defects and persistent luminescence in Eu-doped SrAl$_2$O$_4$}
\author{Khang Hoang}
\email{khang.hoang@ndsu.edu}
\affiliation{Center for Computationally Assisted Science and Technology \& Department of Physics, North Dakota State University, Fargo, North Dakota 58108, United States}

\date{\today}

\begin{abstract}

We investigate native point defects and rare-earth (co)dopants in SrAl$_2$O$_4$ using hybrid density-functional defect calculations. Europium (Eu) and dysprosium (Dy) are found to be mixed valence and energetically most favorable at the Sr lattice sites. However, unlike Eu where both Eu$^{2+}$ and Eu$^{3+}$ can be realized in synthesis, Dy is stable predominantly as Dy$^{3+}$, and the divalent Dy$^{2+}$ may only be photogenerated under irradiation. On the basis of an analysis of Eu-related band--defect (including charge-transfer) and interconfigurational $5d$--$4f$ optical transitions, we assign the characteristic broad blue (445 nm) and green (520 nm) emission bands in Eu$^{2+}$-doped SrAl$_2$O$_4$ to the $4f^65d^1$ $\rightarrow$ $4f^7$ transition in Eu$^{2+}$ incorporated at the Sr1 and Sr2 sites, respectively. Strontium interstitials ({\it not} oxygen vacancies, in contrast to what is commonly believed) and Dy$_{\rm Sr}$ can act as efficient electron traps for room-temperature persistent luminescence. This work calls for a re-assessment of certain assumptions regarding specific carrier trapping centers made in all mechanisms previously proposed for the persistent luminescence in Eu- and (Eu,Dy)-doped SrAl$_2$O$_4$. It also serves as a methodological template for the understanding and design of rare-earth doped phosphors.

\end{abstract}


\maketitle


\section{Introduction}\label{sec;intro}

Persistent luminescence, previously often referred to as phosphorescence or long-lasting phosphorescence, is an intriguing phenomenon in which a material re-emits light over long periods of time after the excitation has stopped \cite{Smet2015Handbook,Xu2019JL}. Persistent luminescent materials (or persistent phosphors) have numerous applications and potential applications in safety signage and toys, road markings, solid-state lighting (flicker reduction), bio-imaging, nighttime solar energy, and photocatalysis \cite{Poelman2020JAP}. Scientific research on persistent luminescence really took off since the discovery of (Eu,Dy)-doped SrAl$_2$O$_4$ by Murayam, Takeuchi, Aoki, and Matsuzawa of Nemoto \& Co., Ltd.~(Japan) in 1993; the material was found to produce extremely bright green and long-lasting (over many hours) luminescence in the dark \cite{Murayam1994patent,Matsuzawa1996JES}. Three decades after the seminal work of the Nemoto researchers, although there has been great progress in understanding the phenomenon and in discovering new persistent phosphors--as it has already been documented in many excellent review articles and book chapters published in the last several years \cite{Smet2015Handbook,Xu2019JL,Poelman2020JAP,RojasHernandez2018RSER,Vitol2019MST,Hagemann2021Handbook}, details of the underlying mechanism for the persistent luminescence observed in rare-earth (RE) doped SrAl$_2$O$_4$ and similar materials are still under debate, and the search for new or improved persistent phosphors, in general, remains largely trial and error.

\begin{figure*}[t]%
\vspace{0.2cm}
\includegraphics*[width=0.98\linewidth]{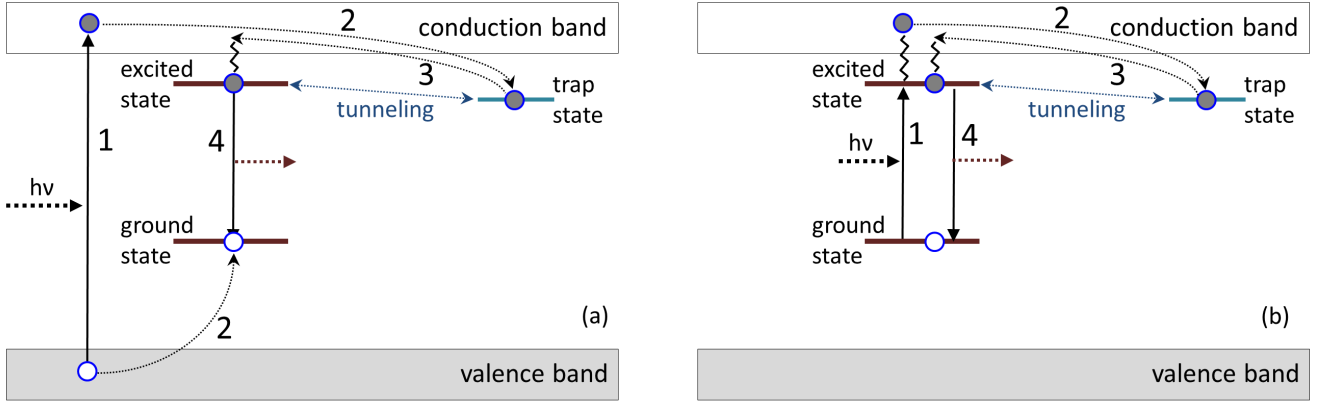}
\caption{Schematic illustration of a persistent luminescence mechanism under (a) band-to-band or (b) below-gap excitation: ({\bf 1}) excitation: electrons are excited and holes generated under illumination; ({\bf 2}) trapping: the excited electrons (generated holes) are captured nonradiatively by electron (hole) traps through the conduction (valence) band and/or via quantum tunneling; ({\bf 3}) detrapping: the trapped electrons are released under thermal stimulation; ({\bf 4}) recombination: the released electrons return to the emission center and recombine with holes, resulting in a delayed luminescence. Adapted from Xu and Tanabe~\cite{Xu2019JL}}
\label{fig;mechanism}
\end{figure*} 

The optical properties of Eu-doped SrAl$_2$O$_4$ are characterized by a broad green emission band peaking at 520 nm at room temperature \cite{Palilla1968JES,Blasse1968PRR,Abbruscato1971ECS}. At lower temperatures, another peak occurs in the emission spectrum at 450 nm (blue) \cite{Poort1995CM}. In addition, a broader excitation band peaking at 276 nm or 250 nm and sharp Eu$^{3+}$ $4f$--$4f$ transitions in the excitation and emission spectra have also been reported \cite{Zollfrank2013AC,Bierwagen2021JL,Botterman2014PRB}. Co-doping the material with Dy does not change the emission spectrum, but strongly enhances the afterglow time and intensity \cite{Matsuzawa1996JES}. There are currently about a dozen different mechanisms proposed in the literature to explain the persistent luminescence observed in Eu- and (Eu,Dy)-doped SrAl$_2$O$_4$ \cite{Matsuzawa1996JES,Beauger1999Thesis,Aitasalo2003JSSC,Aitasalo2004RM,Clabau2005CM,Dorenbos2005JES,Liepina2017JL,Zeng2018JL,Gnidakouong2019CI}. All these mechanisms involve defect levels induced by the RE (co)dopants and/or native point defects. Oxygen vacancies, in particular, have been invoked in many mechanisms as electron trapping centers responsible for the delayed emission. Figure \ref{fig;mechanism} shows a currently generally accepted mechanism involving electron trapping and detrapping processes (a mechanism involving hole trapping and detrapping is similar) \cite{Xu2019JL}. In the context of Eu$^{2+}$-doped SrAl$_2$O$_4$, the emission center is expected to be Eu$^{2+}$ with the ground and excited states being $4f^7$ and $4f^65d^1$, respectively, and the trapping center can be native defects and/or RE co-dopants or other impurities.  

Due to the lack of a detailed understanding of defect physics in SrAl$_2$O$_4$, all the proposed mechanisms \cite{Matsuzawa1996JES,Beauger1999Thesis,Aitasalo2003JSSC,Aitasalo2004RM,Clabau2005CM,Dorenbos2005JES,Liepina2017JL,Zeng2018JL,Gnidakouong2019CI} are highly speculative about specific emission centers and charge carrier trapping centers. This has been a {\it major obstacle} toward rational design of persistent phosphors with improved performance. Here, we present an investigation of native defects and RE (co)dopants in monoclinic SrAl$_2$O$_4$ using first-principles defect calculations. The hybrid density-functional theory (DFT)/Hartree-Fock method \cite{heyd:8207} employed here has been shown to be successful in the study of defects in semiconductors and insulators in general~\cite{Freysoldt2014RMP} and RE-doped materials in particular~\cite{Hoang2015RRL,Hoang2016RRL,Hoang2021PRM}. On the basis of our results, we identify dominant native defects, discuss the stable valence states of the RE (co)dopants, and determine all energy levels induced by the defects. Eu-related optical transitions, including band--defect and interconfigurational Eu$^{2+}$ $5d$--$4f$ absorption and emission processes, are investigated to identify sources of the broad absorption and emission bands observed in experiments, including the characteristic blue and green emissions.

\section{Methodology}\label{sec;method} 

We model defects (i.e., native point defects and impurities) in SrAl$_2$O$_4$ using a supercell approach in which a defect is included in a periodically repeated finite volume of the host material. The formation energy of a defect X in effective charge state $q$ (with respect to the host lattice) is defined as \cite{walle:3851,Freysoldt2014RMP}     
\begin{align}\label{eq:eform}
E^f({\mathrm{X}}^q)&=&E_{\mathrm{tot}}({\mathrm{X}}^q)-E_{\mathrm{tot}}({\mathrm{bulk}}) -\sum_{i}{n_i\mu_i} \\ %
\nonumber &&+~q(E_{\mathrm{v}}+\mu_{e})+ \Delta^q ,
\end{align}
where $E_{\mathrm{tot}}(\mathrm{X}^{q})$ and $E_{\mathrm{tot}}(\mathrm{bulk})$ are the total energies of the defect-containing and perfect bulk supercells, respectively; $n_{i}$ is the number of atoms of species $i$ that have been added ($n_{i}>0$) or removed ($n_{i}<0$) to form the defect; $\mu_{i}$ is the atomic chemical potential, representing the energy of the reservoir with which atoms are being exchanged, and referenced to the total energy per atom of $i$ in its elemental phase at 0 K. $\mu_{e}$ is the chemical potential of electrons, i.e., the Fermi level, representing the energy of the electron reservoir, referenced to the valence-band maximum (VBM) in the bulk ($E_{\mathrm{v}}$). Finally, $\Delta^q$ is the correction term to align the electrostatic potentials of the bulk and defect supercells and to account for finite-size effects on the total energies of charged defects \cite{Freysoldt,Freysoldt11}.

Under thermodynamic equilibrium, the concentration of a defect is directly related to its formation energy \cite{walle:3851}:
\begin{equation}\label{eq;con} 
c=N_{\rm sites}N_{\rm config}\exp{\left(\frac{-E^f}{k_{\rm B}T}\right)}, 
\end{equation} 
where $N_{\rm sites}$ is the number of high-symmetry sites in the lattice (per unit volume) on which the defect can be incorporated, $N_{\rm config}$ is the number of equivalent configurations (per site), and $k_{\rm B}$ is the Boltzmann constant. At a given temperature, a defect with a lower formation energy occurs with a higher concentration. Note, however, that when a material is prepared under non-equilibrium conditions excess defects can be frozen-in and the equilibrium concentration is only the lower bound.

While the Fermi level in Eq.~(\ref{eq:eform}) can be treated as a variable, it is not a free parameter. The actual Fermi-level position of the material can be determined by solving the charge-neutrality equation \cite{walle:3851}:
\begin{equation}\label{eq:neutrality}
\sum_{i}c_{i}q_{i}-n_{e}+n_{h}=0,
\end{equation}
where $c_{i}$ and $q_{i}$ are the concentration and charge, respectively, of defect X$_{i}$; $n_{e}$ and $n_{h}$ are free electron and hole concentrations, respectively; and the summation is over all possible defects present in the material.

The {\it thermodynamic} transition level between charge states $q$ and $q'$ of a defect, $\epsilon(q/q')$, is defined as the Fermi-level position at which the formation energy of the defect in charge state $q$ is equal to that in state $q'$ \cite{Freysoldt2014RMP}, i.e.,
\begin{equation}\label{eq;tl}
\epsilon(q/q') = \frac{E^f(X^{q}; \mu_e=0)-E^f(X^{q'}; \mu_e=0)}{q' - q},
\end{equation}
where $E^f(X^{q}; \mu_e=0)$ is the formation energy of the defect X in charge state $q$ when the Fermi level is at the VBM ($\mu_e=0$). This $\epsilon(q/q')$ level [also referred to as the $(q/q')$ level], corresponding to a defect energy level (or, simply, {\it defect level}), would be observed in experiments where the defect in the final charge state $q'$ fully relaxes to its equilibrium configuration after the transition.

Defect-to-band and band-to-defect optical transitions, including those of the charge-transfer type \cite{Blasse1994Book}, can be characterized using the {\it optical} transition level $E_{\rm opt}^{q/q'}$ that is defined similarly to $\epsilon(q/q')$ but with the total energy of the final state $q'$ calculated using the lattice configuration of the initial state $q$ \cite{Freysoldt2014RMP}. For the $5d$--$4f$ transitions in neutral Eu defects, we calculate the energies based on a constrained occupancy approach and $\Delta$SCF analysis (SCF: self-consistent field), similar to that used in previous studies of RE-doped phosphors \cite{Canning2011PRB,Jia2017PRB}. In this approach, we create the excited state Eu $4f^65d^1$ in SrAl$_2$O$_4$ by manually emptying the highest Eu $4f$ state and filling the next state lying higher in energy. 

The total-energy electronic structure calculations are based on DFT with the Heyd-Scuseria-Ernzerhof (HSE) functional \cite{heyd:8207}, the projector augmented wave (PAW) method \cite{PAW1}, and a plane-wave basis set, as implemented in the Vienna {\it Ab Initio} Simulation Package (\textsc{vasp}) \cite{VASP2}. Along with the CPU version, the GPU port of \textsc{vasp} (version 6.2.1) is also used. The Hartree-Fock mixing parameter is set to 0.33 and the screening length to the default value of 10 {\AA} to match the experimental band gap. We use the PAW potentials in the \textsc{vasp} database which treat Sr $4s^{2}4p^{6}5s^{2}$, Al $3s^{2}3p^{1}$, O $2s^{2}2p^{4}$, Eu $5s^{2}5p^{6}4f^{7}6s^{2}$, and Dy $5s^{2}5p^{6}4f^{10}6s^{2}$ explicitly as valence electrons and the rest as core. Defects are modelled using monoclinic 2$\times$1$\times$3 (168-atom) supercells and integrations over the Brillouin zone are performed using the $\Gamma$ point, except in the constrained occupancy calculations (using \textsc{vasp} 5.3.3) where 1$\times$1$\times$3 (84-atom) supercells and a $\Gamma$-centered 2$\times$2$\times$1 $k$-point mesh are used. In the defect calculations, the lattice parameters are fixed to the calculated bulk values but all the internal coordinates are relaxed. In all the calculations, the plane-wave basis-set cutoff is set to 500 eV and spin polarization is included; structural relaxations are performed with the HSE functional and the force threshold is chosen to be 0.02 eV/{\AA}. Spin-orbit interaction is not included as it has negligible effects on the defect transition levels \cite{Hoang2021PRM}. 

The chemical potentials of Sr, Al, and O vary over a range determined by requiring that the host compound SrAl$_2$O$_4$ is stable against competing Sr--Al--O phases; see Sec.~\ref{sec;results;bulk}. Experimentally relevant or representative sets of $\mu_{\rm Sr}$, $\mu_{\rm Al}$, and $\mu_{\rm O}$ are adopted to present defect formation energies. The chemical potential of Dy is obtained by assuming equilibrium with Dy$_2$O$_3$; that of Eu by assuming equilibrium with Eu$_2$O$_3$ (under oxidizing conditions) or EuO (under reducing conditions). It should be noted that the transition levels $\epsilon(q/q')$ and $E_{\rm opt}^{q/q'}$ are {\it independent} of the choice of the atomic chemical potentials. 

\section{Results and Discussion}\label{sec;results}

\subsection{Bulk properties}\label{sec;results;bulk}

SrAl$_2$O$_4$ crystallizes in the monoclinic $P2_1$ space group \cite{Schulze1981ZAAC}; see Fig.~\ref{fig;unitcell} in Appendix \ref{sec;app}. Its crystal structure has two inequivalent Sr lattice sites, four inequivalent Al sites, and eight inequivalent O sites. The Sr sites, Sr1 and Sr2, are in channels along the $c$-axis formed by the Al--O framework (hereafter referred to as the Sr1 and Sr2 channels). The calculated lattice parameters are $a=8.4491$ {\AA}, $b=8.8159$ {\AA}, $c=5.1525$ {\AA}, and $\beta = 93.5397^\circ$, in excellent agreement with the experimental values \cite{Schulze1981ZAAC}. The calculated band gap is 6.51 eV, an indirect gap with the valence-band maximum (VBM) at the $X$ point in the Brillouin zone and the conduction-band minimum (CBM) at the $\Gamma$ point. For comparison, the reported experimental band gap is 6.5--6.6 eV \cite{Matsuzawa1996JES,Holsa2009JRE,Dutczak2015PCCP}. The VBM consists primarily of the O $2p$ states, whereas the CBM consists of a mixture of the Sr, Al, and O $s$ states. The total static dielectric constant is calculated to be 9.36 (taken as the average of the $xx$, $yy$, and $zz$ components) with the electronic contribution based on the real part of the dielectric function $\epsilon_{1}(\omega)$ for $\omega\rightarrow0$ obtained within HSE and the ionic contribution calculated using density-functional perturbation theory \cite{dielectricmethod} within the generalized-gradient approximation \cite{GGA}.

\begin{figure}
\vspace{0.2cm}
\includegraphics*[width=\linewidth]{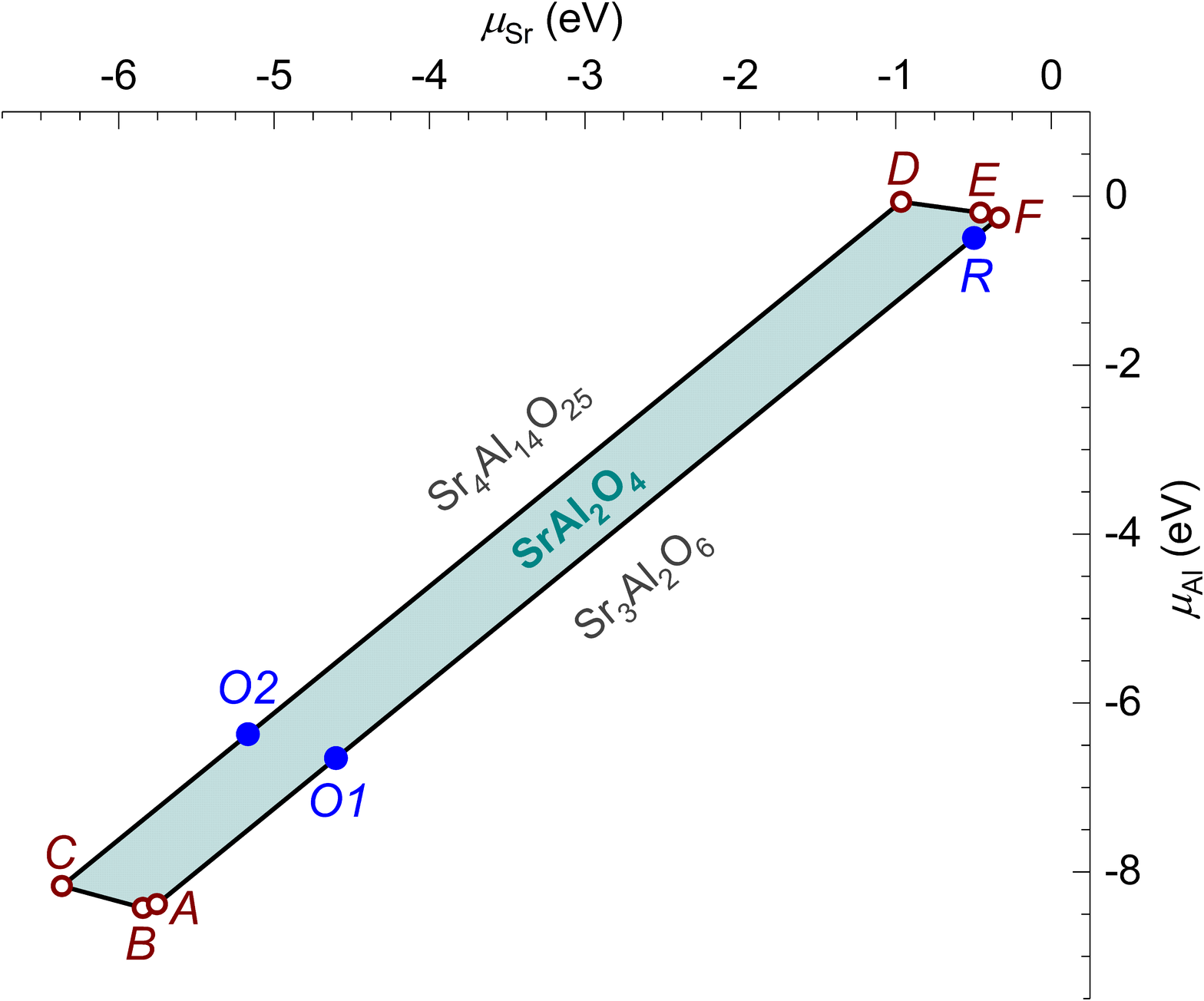}
\caption{Chemical-potential phase diagram for SrAl$_2$O$_4$. Sr--Al--O phases that define the stability region of the host, shown as a shaded polygon, are SrO$_2$ (the $AB$ line), O$_2$ ($BC$), Sr$_4$Al$_{14}$O$_{25}$ ($CD$), SrAl$_4$ ($DE$), SrAl$_2$ ($EF$), and Sr$_3$Al$_2$O$_6$ ($FA$). Points $O1$, $O2$, and $R$ are examples of experimentally relevant oxidizing and reducing environments; see the text.} 
\label{fig;pd} 
\end{figure}

Figure \ref{fig;pd} shows the phase stability of SrAl$_2$O$_4$. The formation enthalpies (calculated at 0 K) of different Sr--Al--O phases are listed in Table \ref{tab;enthalpies} in Appendix \ref{sec;app}. The initial structures of these phases are taken from the Materials Project database \cite{MPdatabase}. The stability region of SrAl$_2$O$_4$ is delineated mainly by Sr$_3$Al$_2$O$_6$ and Sr$_4$Al$_{14}$O$_{25}$, which is consistent with the experimental phase diagram \cite{Massazza1959,Capron2002JACerS}. 

\begin{figure*}
\vspace{0.2cm}
\includegraphics*[width=\linewidth]{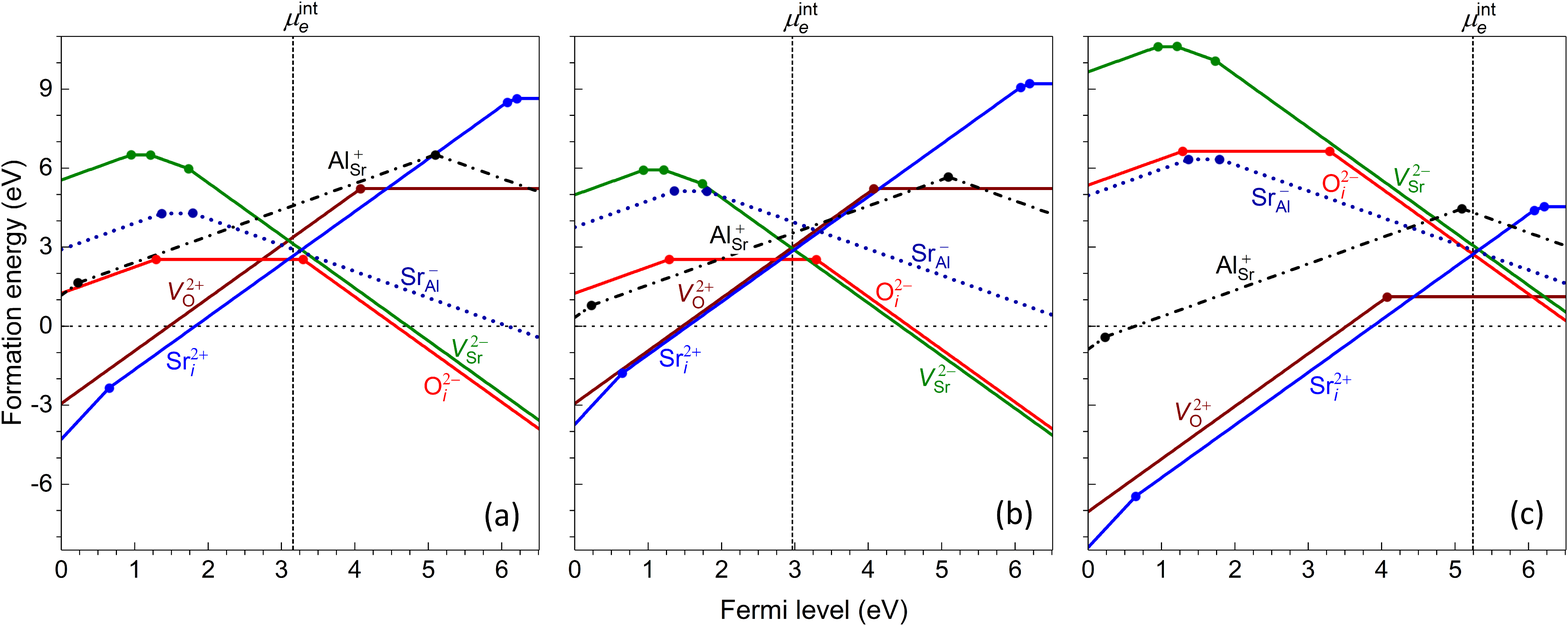}
\caption{Formation energies of native defects in SrAl$_2$O$_4$, as a function of the Fermi level from the VBM (0 eV) to the CBM (6.51 eV), calculated at points (a) $O1$, (b) $O2$, and (c) $R$ in the phase diagram (Fig.~\ref{fig;pd}). For each defect, only segments of the formation energy lines corresponding to the lowest-energy charge states are shown. The slope of these segments indicates the charge state $q$: positively (negatively) charged defect configurations have positive (negative) slopes; horizontal segments correspond to neutral defect conﬁgurations. Large dots connecting two segments with different slopes mark the {\it defect levels} $\epsilon(q/q')$. For a defect with multiple inequivalent lattice sites, only the lowest-energy lattice site is reported.} 
\label{fig;fe;native} 
\end{figure*}

In the presentation of defect formation energies in the next sections, we make use of the following points in the phase diagram: (i) $O1$, where the host compound is in equilibrium with Sr$_3$Al$_2$O$_6$ and air (the oxygen partial pressure $p_{{\rm O}_2} = 0.21$ atm) at 750$^\circ$C, (ii) $O2$, where the host is in equilibrium with Sr$_4$Al$_{14}$O$_{25}$ and air at 750$^\circ$C, and (iii) $R$, where the host is in equilibrium with Sr$_3$Al$_2$O$_6$ and Ar/H$_2$ 5\%  ($p_{{\rm O}_2} \sim 10^{-20}$ atm) at 1400$^\circ$C. Points $O1$ and $O2$ (with $\mu_{\rm O} = -1.20$ eV) represent oxidizing environments, whereas point $R$ ($\mu_{\rm O} = -5.30$ eV) represents a highly reducing environment. Here, $\mu_{\rm O}$ is calculated as half of the Gibbs free energy of O$_2$ gas at the given $T$ and $p_{{\rm O}_2}$ values \cite{stull1971}. These conditions are chosen to reflect the conditions under which undoped and Eu-doped SrAl$_2$O$_4$ samples are often prepared. Beauger \cite{Beauger1999Thesis} reported the presence of Sr$_3$Al$_2$O$_6$ as an impurity phase, indicating that their synthesis environment corresponds to a point very close to the $FA$ line in Fig.~\ref{fig;pd}.    

\subsection{Native point defects}\label{sec;results;natives}

\begin{figure*}
\vspace{0.2cm}
\includegraphics*[width=0.85\linewidth]{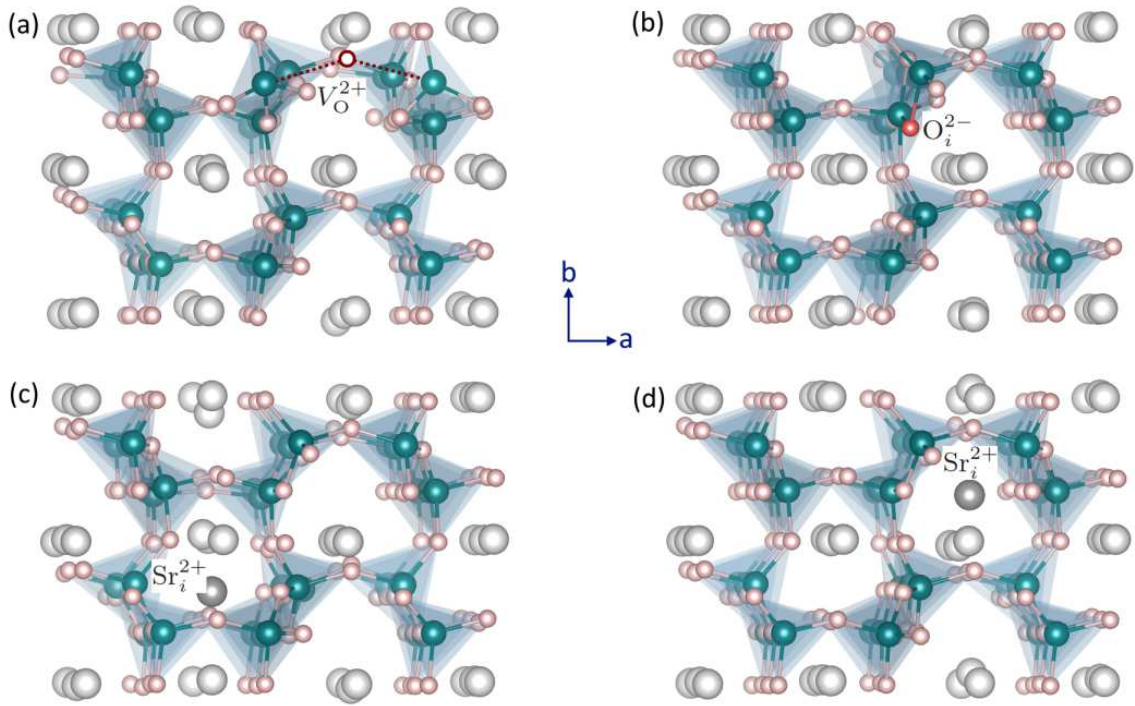}
\caption{Structures of selected native defect configurations in SrAl$_2$O$_4$: (a) $V_{\rm O}^{2+}$ at the O8 lattice site, (b) O$_i^{2-}$, (c) Sr$_i^{2+}$ in the Sr1 channel, and (d) Sr$_i^{2+}$ in the Sr2 channel. Large (gray) spheres are Sr, medium (blue) are Al, and small (red) are O. All the atomic structures in this work are visualized using the \textsc{vesta} package \cite{VESTA}.} 
\label{fig;native;struct} 
\end{figure*}

Figure \ref{fig;fe;native} shows the formation energy of native defects in SrAl$_2$O$_4$. The defects introduce one or more energy levels in the host's band gap region (marked by large dots in the figure; explicit numerical values are listed in Table \ref{tab;defectlevel}). Under the experimentally relevant conditions (see Sec.~\ref{sec;results;bulk}), the dominant defects (i.e., those with the lowest formation energy) are Sr and O vacancies and interstitials. In the absence of impurities, intentionally doped or unintentionally present, the Fermi level is ``pinned'' at the position $\mu_e^{\rm int}$ (``int'' for intrinsic), determined predominantly by the lowest-energy charged defects, where the charge neutrality condition (\ref{eq:neutrality}) is maintained. From one point in the phase diagram (Fig.~\ref{fig;pd}) to another, the {\it defect landscape} in SrAl$_2$O$_4$ changes, leading to a change in the $\mu_e^{\rm int}$ value; e.g., $\mu_e^{\rm int}$ moves toward the CBM in going from oxidizing (points $O1$ and $O2$) to reducing (point $R$) synthesis conditions. We find that $\mu_e^{\rm int}$ varies from 2.28 eV (under the conditions at point $C$) to 5.36 eV (point $F$). The Fermi level of SrAl$_2$O$_4$, even in the presence of intentional and/or unintentional impurities, cannot be close to the VBM (under all synthesis conditions) but can, in principle, be high up at the CBM (under reducing synthesis conditions only). This is because the formation energy of certain native defects, specifically Sr interstitials and O vacancies, is negative in the region near the VBM (up to $E_{\mathrm{v}} + 0.94$ eV at point $C$ and higher at other points in the phase diagram), whereas the native defects can all have a positive formation energy under the reducing conditions (e.g., those at points $D$--$F$ and $R$).

The removal of an O$^{2-}$ results in a positively charged O vacancy ($V_{\rm O}^{2+}$, spin $S=0$). Other charge states, $V_{\rm O}^+$ ($S=1/2$) and $V_{\rm O}^{0}$ ($S=0$), are also stable. The energetics and electronic behavior of $V_{\rm O}$ are different at the inequivalent O lattice sites due to the slightly different local lattice environments. At certain O sites (e.g., at the O8 site as shown in Fig.~\ref{fig;fe;native}), $V_{\rm O}^+$ is energetically less favorable than $V_{\rm O}^{2+}$ and $V_{\rm O}^0$ in the entire range of the Fermi-level values. Notably, the energy levels introduced by the vacancy at the O1 to O8 sites are all about 2.43--3.55 eV below the CBM; see Fig.~\ref{fig;vo}. The lowest-energy $V_{\rm O}^{2+}$ configuration occurs at the O8 site, see Fig.~\ref{fig;native;struct}(a), indicating that the Al--O8 bonds are weakest. The highest-energy $V_{\rm O}^{2+}$ occurs at the O1 site. Under reducing conditions (e.g., at point $R$), $V_{\rm O}$ occurs in the form of $V_{\rm O}^0$ with a high concentration (highest among all native defects).

Our results for the oxygen vacancies are thus in {\it sharp contrast} to those of Finley et al.~\cite{Finley2018JPCC} where the defect energy levels introduced by $V_{\rm O}$ were found to scatter all over the upper half of the host band gap region. Given the similarity in the bonding environments of the inequivalent O sites (e.g., every single O atom is bonded to two Al atoms), the almost random distribution of the defect levels reported in Ref.~\cite{Finley2018JPCC} cannot be justified.

The addition of an oxygen, which results in an oxygen interstitial (O$_i$), introduces energy levels near the midgap region; see also Table \ref{tab;defectlevel}. In the O$_i^{2-}$ ($S=0$) configuration, the added O$^{2-}$ is shared between two AlO$_{4}$ units; Fig.~\ref{fig;native;struct}(b) [and Fig.~\ref{fig;native;struct;extra}(a)]. O$_i^{-}$ ($S=1/2$; not visible in Fig.~\ref{fig;fe;native} due to its very small stability range) is a complex of O$_i^{2-}$ and an electron hole localized at an O site (hereafter referred to as $\eta_{\rm O}^+$, $S=1/2$). The structure of O$_i^0$ [$S=0$; Fig.~\ref{fig;native;struct;extra}(b)] can be described as having the interstitial oxygen bonded to one of the O atoms in an AlO$_4$ unit (the O--O distance is 1.48 {\AA}), resulting in a distorted AlO$_5$. O$_i^+$ ($S=1/2$) is a complex of O$_i^0$ and $\eta_{\rm O}^+$.       

The removal of a Sr$^{2+}$ ion from the host lattice results in a negatively charged Sr vacancy ($V_{\rm Sr}^{2-}$, $S=0$). Other stable charge states of $V_{\rm Sr}$ are $V_{\rm Sr}^{-}$ ($S=1/2$), $V_{\rm Sr}^{0}$ ($S=1$), and $V_{\rm Sr}^{+}$ ($S=3/2$), which can be regarded as complexes consisting of $V_{\rm Sr}^{2-}$ and one, two, and three $\eta_{\rm O}^+$ defects, respectively. $V_{\rm Sr}$ at the Sr1 site is slightly lower in energy than that at the Sr2 site, e.g., by 0.08 eV in the case of $V_{\rm Sr}^{2-}$. The defect levels associated with $V_{\rm Sr}$  are in the lower half of the band gap region; see also Table \ref{tab;defectlevel}.

There are two possible sites for Sr interstitials (Sr$_i$): one, Sr$_{i1}$, locating in the Sr1 channel (along the $c$-axis) but off the Sr1 chain and between two Al atoms (along the $a$-axis), see Fig.~\ref{fig;native;struct}(c), and the other, Sr$_{i2}$, in the Sr2 channel but off the Sr2 chain and between two Al atoms (along the $a$-axis), see Fig.~\ref{fig;native;struct}(d) [and Fig.~\ref{fig;native;struct;extra}(c)]. Sr$_{i1}$ and Sr$_{i2}$ are approximately at the interstitial sites $V2$ and $V1$, respectively, mentioned in Bierwagen et al.~\cite{Bierwagen2021JL}. Sr$_{i2}$ is lower in energy than Sr$_{i1}$ (e.g., by 0.19 eV in the case of Sr$_i^{2+}$, $S=0$). The defect introduces three defect levels: one above the VBM and two just below the CBM. The other charge states are Sr$_i^{3+}$ ($S=1/2$; a complex of Sr$_i^{2+}$ and $\eta_{\rm O}^+$), Sr$_i^{+}$  ($S=1/2$), and Sr$_i^{0}$ ($S=0$). At $\mu_e^{\rm int}$, Sr$_i^{2+}$ is energetically most stable among the stable charge states of Sr$_i$ and one of the lowest-energy native defects.   

Antisite defects, Al$_{\rm Sr}$ and Sr$_{\rm Al}$, are also considered. We find that they are higher in energy than the Sr and O vacancies and interstitials. Al$_{\rm Sr}$ is lower in energy at the Sr2 site than at the Sr1 site, e.g., 0.55 eV lower in the case of Al$_{\rm Sr}^+$ ($S=0$); Al$_{\rm Sr}^{2+}$ ($S=1/2$) is a complex of Al$_{\rm Sr}^+$ and $\eta_{\rm O}^+$. Sr$_{\rm Al}$ is energetically most favorable at the Al2 site; Sr$_{\rm Al}^0$ ($S=1/2$) and Sr$_{\rm Al}^+$ ($S=1$) are defect complexes of Sr$_{\rm Al}^-$ and one and two $\eta_{\rm O}^+$, respectively. Finally, the creation of Al vacancies ($V_{\rm Al}$) involves breaking four strongly covalent  Al--O bonds, a high energy process. Such defects, as well as Al interstitials (Al$_i$), have high formation energies and are thus not included here.    

\begin{figure*}
\vspace{0.2cm}
\includegraphics*[width=\linewidth]{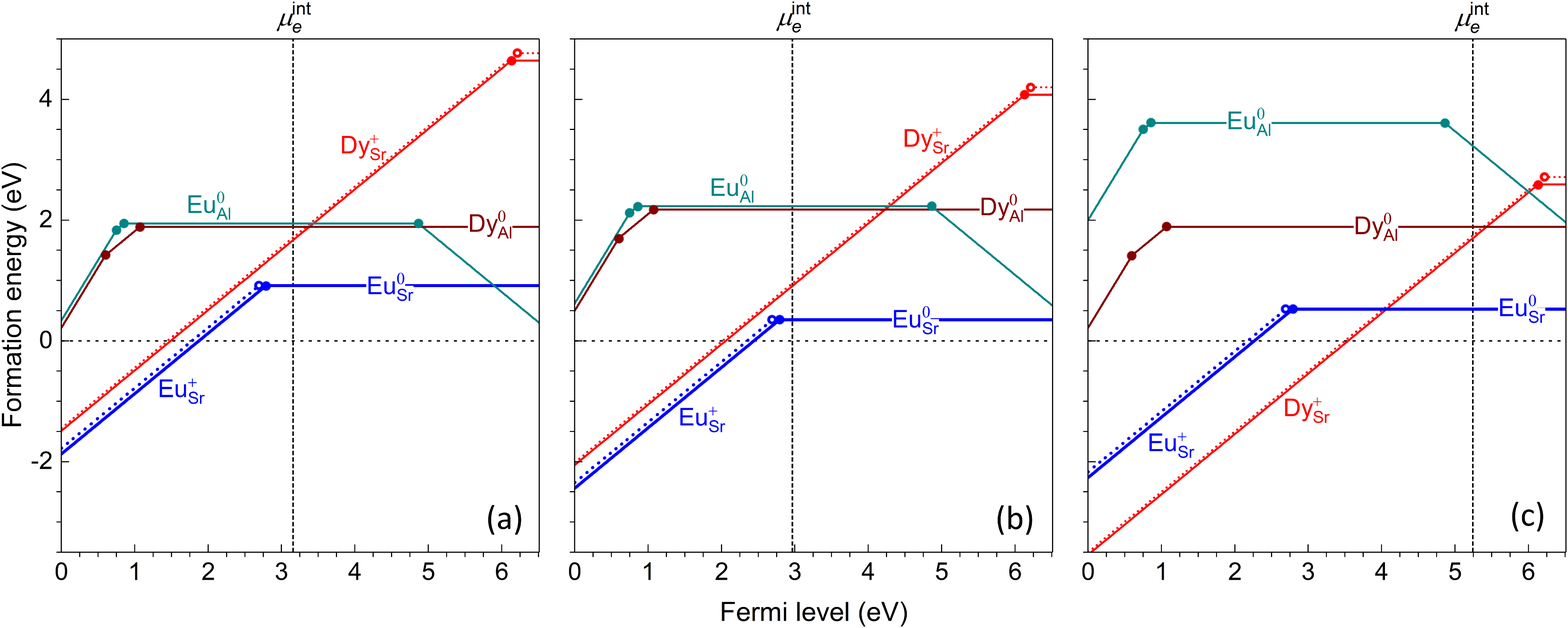}
\caption{Formation energies of Eu- and Dy-related defects in SrAl$_2$O$_4$, calculated at points (a) $O1$, (b) $O2$, and (c) $R$ in the phase diagram (Fig.~\ref{fig;pd}). Large dots connecting two segments with different slopes mark the {\it defect levels}. For Eu$_{\rm Sr}$ and Dy$_{\rm Sr}$, the defect configurations at both the Sr1 (dotted lines) and Sr2 (solid lines) sites are included.} 
\label{fig;fe;re} 
\end{figure*}

\subsection{Rare-earth (co)dopants}\label{sec;results;re}

Figure \ref{fig;fe;re} shows the formation energy of substitutional Eu and Dy impurities at the Sr and Al sites. Results for the REs at the interstitial sites are included in Fig.~\ref{fig;re;all}. Table \ref{tab;defectlevel} lists explicit numerical values of the energy levels and stable RE ions. We find that, at and near the Fermi level determined by the native defects ($\mu_e^{\rm int }$), RE$_{\rm Sr}$ is lower in energy than RE$_{\rm Al}$ and RE$_i$, indicating that Eu and Dy are incorporated into SrAl$_2$O$_4$ at the Sr sites. 

\begin{figure}
\centering
\includegraphics[width=\linewidth]{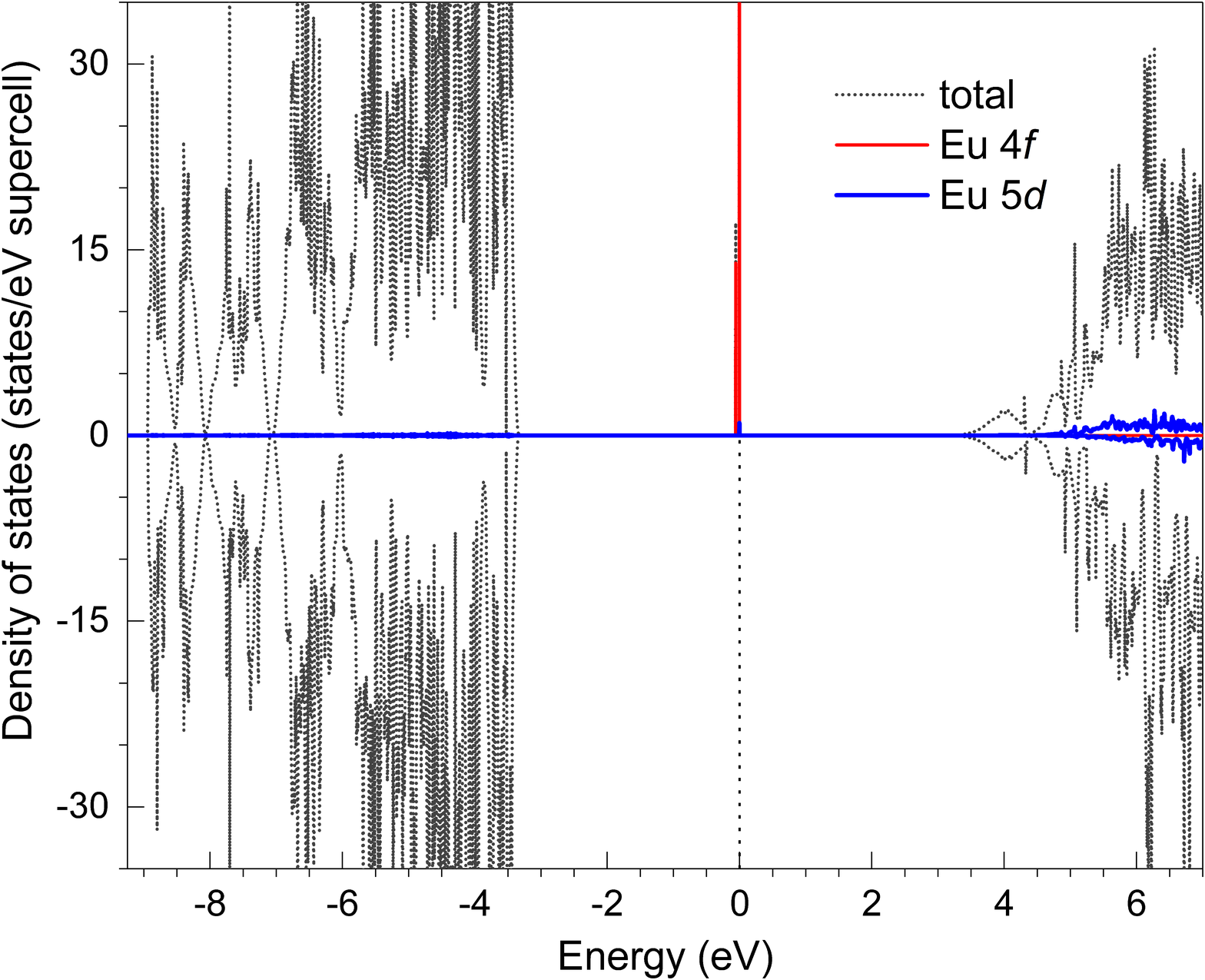}
\caption{Total and projected densities of states of Eu-doped SrAl$_2$O$_4$, specifically the Eu$_{\rm Sr}^0$ defect configuration with the chemical composition Eu$_x$Sr$_{1-x}$Al$_2$O$_4$ ($x=0.125$). The zero of energy is set to the highest occupied state.}
\label{fig;dos;eu}
\end{figure}

Eu$_{\rm Sr}$ is stable as Eu$_{\rm Sr}^0$ (i.e., Eu$^{2+}$, with a magnetic moment of 7$\mu_{\rm B}$; spin $S=7/2$) and/or Eu$_{\rm Sr}^+$ (i.e., Eu$^{3+}$, with a magnetic moment of 6$\mu_{\rm B}$; $S=3$). The Eu$_{\rm Sr}^0$ configurations at the Sr1 and Sr2 sites are almost degenerate in energy, whereas Eu$_{\rm Sr}^+$ at the Sr2 site is 0.09 eV lower in energy than at the Sr1 site. The $(+/0)$ level of Eu$_{\rm Sr}$ is 2.69 eV above the VBM when incorporated at the Sr1 site or 2.79 eV when incorporated at the Sr2 site; below (above) this level, Eu$^{3+}$ (Eu$^{2+}$) is energetically more favorable. The Eu$^{2+}$/Eu$^{3+}$ ratio thus depends on the actual position of the Fermi level which, in turn, depends on the synthesis conditions. The ratio is high under reducing conditions (e.g., at point $R$) and decreases as one changes from reducing to oxidizing conditions. Note that, under actual synthesis conditions and as Eu-doped SrAl$_2$O$_4$ is typically prepared using Eu$^{3+}$ as dopant, an equilibrium assumption may not hold true, and the trivalent ion may be frozen in \cite{Bierwagen2021JL}. In other words, Eu$^{3+}$ may be present even in samples prepared under less oxidizing conditions. 

The mixed valence of Eu in SrAl$_2$O$_4$ can be understood based on the calculated electronic structure of Eu$_{\rm Sr}^0$, reported in Fig.~\ref{fig;dos;eu}. The HSE calculations of the electronic density of states are carried out using a smaller, $1\times1\times2$ supercell and a $\Gamma$-centered $3\times3\times3$ $k$-point mesh. Eu$_{\rm Sr}^0$ has seven occupied spin-up $4f$ states in the host band gap (and seven spin-down unoccupied $4f$ states deep in the conduction band). Given the electronic structure, when one electron is removed from this neutral defect configuration, the electron is removed from the highest occupied state (which is the highest Eu $4f$ state). This results in the divalent Eu$^{2+}$ ($4f^7$) being oxidized to the trivalent Eu$^{3+}$ ($4f^6$), i.e., Eu$_{\rm Sr}^0$ to Eu$_{\rm Sr}^+$. Figure~\ref{fig;chgdiff}(a) shows the localized Eu $4f$ electron associated with Eu$_{\rm Sr}^0$. 

At and near $\mu_e^{\rm int}$, Eu$_{\rm Al}$ is stable as Eu$_{\rm Al}^0$ (i.e., Eu$^{3+}$) under the oxidizing conditions ($O1$ and $O2$), and as Eu$_{\rm Al}^-$ (i.e., Eu$^{2+}$) and/or Eu$_{\rm Al}^0$ under the reducing condition ($R$). Other electronically stable charge states are Eu$_{\rm Al}^+$ (a complex of Eu$_{\rm Al}^0$ and $\eta_{\rm O}^+$) and Eu$_{\rm Al}^{2+}$ (a complex of Eu$_{\rm Al}^0$ and two $\eta_{\rm O}^+$). These positively charged states, however, cannot be obtained during synthesis due to the negative formation energies of Eu$_{\rm Sr}$ and the native defects in the region above the VBM. We find that, at and near $\mu_e^{\rm int}$, Eu$_i$ is electronically stable as Eu$_i^{2+}$ (i.e., Eu$^{2+}$) and has a much higher formation energy than Eu$_{\rm Sr}$ and Eu$_{\rm Al}$. 

The mixed valence of Eu is well discussed in the experimental literature \cite{Beauger1999Thesis,Clabau2005CM,Behrh2015AC,Bierwagen2021JL}. The fact that both Eu$^{3+}$ and Eu$^{2+}$ can be realized in as-prepared SrAl$_2$O$_4$ is consistent with the above results showing the $(+/0)$ level of Eu$_{\rm Sr}$ is located near midgap and the stability ranges of Eu$_{\rm Sr}^+$ and Eu$_{\rm Sr}^0$ are accessible during synthesis. Wang et al.~\cite{Wang2002JL} found Eu$^{2+}$ to be distributed almost equally at the two Sr sites, consistent with the fact that the energies of Eu$_{\rm Sr}^0$ at the Sr1 and Sr2 sites are almost equal. Note that other valence states of Eu (e.g., Eu$^+$, as proposed by Matsuzawa et al.~\cite{Matsuzawa1996JES}) cannot be stabilized electronically.

\begin{figure}
\centering
\includegraphics[width=\linewidth]{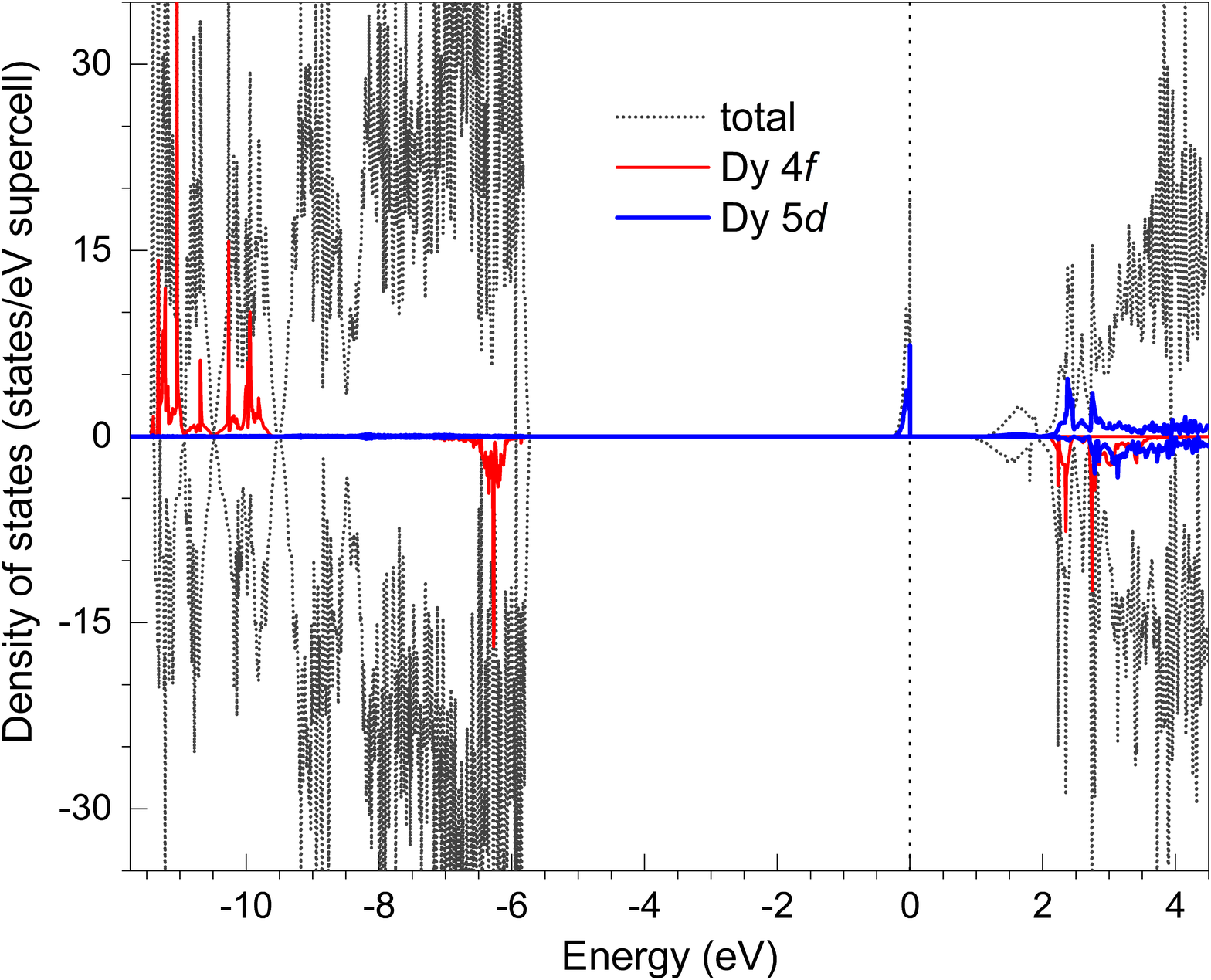}
\caption{Total and projected densities of states of Dy-doped SrAl$_2$O$_4$, specifically the Dy$_{\rm Sr}^0$ defect configuration with the chemical composition Dy$_x$Sr$_{1-x}$Al$_2$O$_4$ ($x=0.125$). The zero of energy is set to the highest occupied state.}
\label{fig;dos;dy}
\end{figure}

Dy$_{\rm Sr}$ is energetically favorable as Dy$_{\rm Sr}^+$ (i.e., Dy$^{3+}$; $S=5/2$) in almost the entire range of the Fermi-level values and as Dy$_{\rm Sr}^0$ (i.e., Dy$^{2+}$; $S=3$) in a small range below the CBM. The $(+/0)$ level is located at 6.21 eV (6.13 eV) above the VBM, i.e., 0.30 eV (0.38 eV) from the CBM, at the Sr1 (Sr2) site. The defect is slightly lower in energy at the Sr2 site than at the Sr1 site; the difference is 0.12 eV for Dy$_{\rm Sr}^0$ or 0.05 eV for Dy$_{\rm Sr}^+$. We find that the electronic configuration of Dy$^{3+}$ is $4f^9$, whereas that of Dy$^{2+}$ is $4f^95d^1$. Figure \ref{fig;dos;dy} shows the electronic structure of Dy$_{\rm Sr}^0$. The nine (seven spin-up and two spin-down) occupied Dy $4f$ states are in the valence band, the occupied Dy $5d^1$ state is in the upper half of the host band gap, and the five spin-down unoccupied Dy $4f$ states are in the conduction band. When one electron is removed from Dy$_{\rm Sr}^0$, it is removed from the Dy $5d^1$ state, which leads to the formation of Dy$_{\rm Sr}^+$. Figure~\ref{fig;chgdiff}(c) shows the localized Dy $5d$ electron associated with the Dy$_{\rm Sr}^0$ configuration.

Dy$_{\rm Al}$ is stable as Dy$_{\rm Al}^0$ (i.e., Dy$^{3+}$), except near the VBM where it is stable as Dy$_{\rm Al}^+$ (i.e., Dy$^{4+}$ with the electronic configuration $4f^8$) or Dy$_{\rm Al}^{2+}$ (a complex of Dy$_{\rm Al}^+$ and $\eta_{\rm O}^+$). However, given the negative formation energy of Dy$_{\rm Sr}$ and the native defects in the region above the VBM, these positive charge states are inaccessible during synthesis. Dy$_i$ is stable as Dy$_i^{3+}$; see Fig.~\ref{fig;re;all}. At and near $\mu_e^{\rm int}$, Dy$_{\rm Al}$ and Dy$_i$ are all higher energy than Dy$_{\rm Sr}$, indicating that Dy is incorporated at the Sr sites.

The results for the Dy-related defects thus confirm the stabilization of Dy$_{\rm Sr}^+$, consistent with the fact that Dy is present as Dy$^{3+}$ in as-prepared Dy-doped samples. Interestingly, Dy$_{\rm Sr}^0$ (i.e., Dy$^{2+}$) is also found to be structurally and electronically stable. Dy$_{\rm Sr}^0$, however, has a very small stability range that is close to the CBM, and in that range Dy$_{\rm Sr}$ becomes much less favorable energetically than Dy$_{\rm Al}^0$ (i.e., Dy$^{3+}$), indicating that the divalent Dy$^{2+}$ is almost impossible to obtain during synthesis. It can be photogenerated under irradiation nonetheless. Experimentally, Joos et al.~\cite{Joos2020PRL} reported evidence of the Dy$^{3+/2+}$ valence change in (Eu, Dy)-doped Sr$_4$Al$_{14}$O$_{25}$ under laser excitation. A similar process for Dy could be observed in SrAl$_2$O$_4$. Note that Dorenbos \cite{Dorenbos2005JES} estimated the ``ground state of Dy$^{2+}$'' to be at 0.9 eV below the CBM, based on a semiempirical model. Such a level is much lower than our calculated $(+/0)$ level of Dy$_{\rm Sr}$. Our results also show that Dy$^{4+}$ is not stable electronically at the Sr site; the tetravalent ion is stable at the Al site but energetically unfavorable, as discussed above. 

\begin{figure*}
\vspace{0.2cm}
\includegraphics*[width=\linewidth]{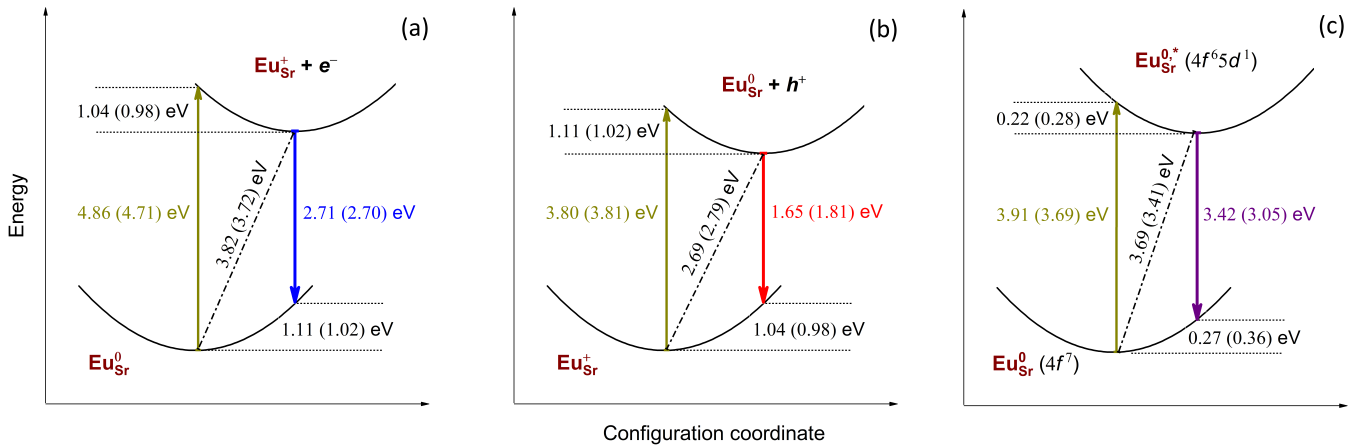}
\caption{Configuration-coordinate diagram illustrating optical absorption (up arrow) and emission (down arrow) processes for Eu$_{\rm Sr}$ in SrAl$_2$O$_4$: band--defect transitions involving the $(+/0)$ defect level of Eu$_{\rm Sr}$ exchanging (a) electrons with the CBM or (b) holes with the VBM, and (c) interconfigurational $5d$--$4f$ transitions in Eu$_{\rm Sr}^0$. The dash-dotted line indicates the thermal energy (i.e., ZPL). The values sandwiched between two dotted lines are the relaxation energies (i.e., the Franck-Condon shifts). The values outside (inside) the brackets are for Eu at the Sr1 (Sr2) lattice site. Axes are not to scale.} 
\label{fig;fe;cc} 
\end{figure*}

Finally, we consider possible association between Eu$_{\rm Sr}$ and Dy$_{\rm Sr}$ or a native defect. Figure \ref{fig;complex} shows the formation energy of Eu$_{\rm Sr}$-Dy$_{\rm Sr}$, Eu$_{\rm Sr}$-$V_{\rm O}$, Eu$_{\rm Sr}$-O$_i$, Eu$_{\rm Sr}$-$V_{\rm Sr}$, and Eu$_{\rm Sr}$-Sr$_i$; see also Table \ref{tab;complex} for details on the stable charge states of the complexes. Focusing on the Fermi-level range near the CBM, which is relevant to the physics under investigation, we find that the electronic behavior of the complexes is determined predominantly by the non-Eu constituent. Specifically, (Eu$_{\rm Sr}$-Dy$_{\rm Sr}$)$^{+}$ is a defect complex consisting of Eu$_{\rm Sr}^0$ and Dy$_{\rm Sr}^+$ and (Eu$_{\rm Sr}$-Dy$_{\rm Sr}$)$^{0}$ is a complex of Eu$_{\rm Sr}^0$ and Dy$_{\rm Sr}^0$; (Eu$_{\rm Sr}$-$V_{\rm O}$)$^{0}$ is a complex of Eu$_{\rm Sr}^0$ and $V_{\rm O}^{0}$; (Eu$_{\rm Sr}$-O$_i$)$^{2-}$ is a complex of Eu$_{\rm Sr}^0$ and O$_i^{2-}$; (Eu$_{\rm Sr}$-$V_{\rm Sr}$)$^{2-}$ is a complex of Eu$_{\rm Sr}^0$ and $V_{\rm Sr}^{2-}$; (Eu$_{\rm Sr}$-Sr$_i$)$^{2+}$ is a complex of Eu$_{\rm Sr}^0$ and Sr$_i^{2+}$, (Eu$_{\rm Sr}$-Sr$_i$)$^{+}$ is a complex of Eu$_{\rm Sr}^0$ and Sr$_i^{+}$, and (Eu$_{\rm Sr}$-Sr$_i$)$^{0}$ is a complex of Eu$_{\rm Sr}^0$ and Sr$_i^{0}$. Note that we determine the charge state of a defect configuration by examining the calculated total and local magnetic moments, electron occupation, and local lattice environment. The $(+/0)$ level of Eu$_{\rm Sr}$-Dy$_{\rm Sr}$ is 6.11 eV above the VBM, almost the same as that of the isolated Dy$_{\rm Sr}$; the $(2+/+)$ and $(+/0)$ levels of Eu$_{\rm Sr}$-Sr$_i$ are 6.04 eV and 6.21 eV above the VBM, almost the same as those of the isolated Sr$_i$. 

Most notably, we find that the binding energy of the Eu-related complexes is very small or even negative, see Table \ref{tab;complex}, indicating that they are not stable under thermodynamic equilibrium, which is also consistent with the above analysis of the electronic behavior. In other words, Eu$_{\rm Sr}$ is unlikely to stay close and form a defect complex with Dy$_{\rm Sr}$ or any of the dominant native defects in SrAl$_2$O$_4$, and even if it does, e.g., when the constituent defects get trapped next to each other, the electronic behavior of the complex in the relevant range of Fermi-level values is not that different from the isolated constituents. Our results for Eu$_{\rm Sr}$-Dy$_{\rm Sr}$ are thus consistent with experimental observations that, in (Eu$^{2+}$,RE$^{3+}$)-doped SrAl$_2$O$_4$, the RE$^{3+}$ co-dopant only enhances the afterglow time and intensity and does not change the position or the shape of the emission band \cite{Matsuzawa1996JES,Katsumata2006JACerS}.

\subsection{Eu-related optical transitions}\label{sec;results;lumin}

Let us now examine possible band--defect and interconfigurational $5d$--$4f$ optical transitions involving the Eu$_{\rm Sr}$ defect (The intraconfigurational Eu $4f$--$4f$ optical transitions are not explicitly considered in this work).

Figure \ref{fig;fe;cc}(a) illustrates the optical absorption (defect-to-band) and emission (band-to-defect) processes involving the $(+/0)$ level of Eu$_{\rm Sr}$ exchanging electrons with the CBM. Under illumination, for example, the Eu$_{\rm Sr}^0$ configuration (e.g., present in as-prepared Eu-doped SrAl$_2$O$_4$ samples) can absorb a photon and become ionized to Eu$_{\rm Sr}^+$ with the removed electron being excited into the conduction band. The peak absorption energy ($E_{\rm abs}$) related to the optical transition level $E_{\rm opt}^{0/+}$ (the formation energy difference between Eu$_{\rm Sr}^0$ and the Eu$_{\rm Sr}^+$ configuration in the lattice geometry of Eu$_{\rm Sr}^0$) is 4.86 eV (4.71 eV), with a relaxation energy (i.e., the Franck-Condon shift, $d_{\rm FC}^{\rm e}$) of 1.04 eV (0.98 eV), when Eu is incorporated at the Sr1 (Sr2) lattice site. With such a large relaxation energy, the emission is expected to be broad. In the reverse process, Eu$_{\rm Sr}^+$ can capture an electron from the CBM, e.g., previously excited from Eu$_{\rm Sr}^0$ (or from the valence band) to the conduction band. {\it If} the recombination is {\it radiative}, a photon will be emitted. The peak emission energy ($E_{\rm em}$) related to the optical transition level $E_{\rm opt}^{+/0}$ (the formation energy difference between Eu$_{\rm Sr}^+$ and the Eu$_{\rm Sr}^0$ configuration in the lattice geometry of Eu$_{\rm Sr}^+$) is 2.71 eV (2.70 eV), with a relaxation energy ($d_{\rm FC}^{\rm g}$) of 1.11 eV (1.02 eV), at the Sr1 (Sr2) site. The thermal energy ($E_{\rm therm}$; also referred to as the zero-phonon line or ZPL) associated with the Eu$_{\rm Sr}^0$ $\rightleftharpoons$ Eu$_{\rm Sr}^+$ + $e^-$ transitions is 3.38 eV (3.25 eV) at the Sr1 (Sr2) site, measured from the CBM. The ZPL marks the initial {\it onset} of the absorption band.

Note, however, that the band-to-defect emission process Eu$_{\rm Sr}^+$ + $e^-$ $\rightarrow$ Eu$_{\rm Sr}^0$ may not occur at all due to another, competing process in which the recombination is {\it nonradiatively}: Eu$_{\rm Sr}^+$ + $e^-$ $\rightarrow$ Eu$_{\rm Sr}^{0,\ast}$, where Eu$_{\rm Sr}^{0,\ast}$ is Eu$^{2+}$ in its excited state $4f^65d^1$. In this case, Eu$_{\rm Sr}^{0,\ast}$ will relax to its ground state Eu$_{\rm Sr}^0$ and release a photon through the allowed $5d$--$4f$ transition (discussed below).

In addition to exchanging electrons with the CBM, the $(+/0)$ level of Eu$_{\rm Sr}$ can also exchange holes with the VBM. Figure \ref{fig;fe;cc}(b) illustrates the absorption and emission Eu$_{\rm Sr}^+$ $\rightleftharpoons$ Eu$_{\rm Sr}^0$ + $h^+$ processes. In the literature, these processes are often regarded as involving an O$^{2-}$ to Eu$^{3+}$ charge transfer (CT) and referred to as CT processes \cite{Clabau2005CM,Zollfrank2013AC}. The hole ($h^+$) state at the VBM consists primarily of the O $2p$ states, as discussed earlier. Our calculations show a different set of the absorption, emission, relaxation, and thermal energies as indicated in Fig.~\ref{fig;fe;cc}(b). Similar to the earlier case, the band-to-defect emission Eu$_{\rm Sr}^0$ + $h^+$ $\rightarrow$ Eu$_{\rm Sr}^+$ may not be observed. This is because the energy from the recombination of the electron localized at Eu$_{\rm Sr}^0$ and the free hole can quickly be absorbed into the $4f$-electron core of Eu$^{3+}$ which then excites the ion and leads to intra-$f$ luminescence, as previously discussed in the case of RE-doped GaN \cite{Hoang2022PRM}. 

Experimentally, Botterman et al.~\cite{Botterman2014PRB} reported a broad excitation band peaking at 276 nm (4.49 eV); Zollfrank et al.~\cite{Zollfrank2013AC} and Bierwagen et al.~\cite{Bierwagen2021JL} reported a similar value, at 250 nm (4.96 eV). This band has often been assigned to an O$^{2-}$ to Eu$^{3+}$ charge transfer \cite{Zollfrank2013AC}, which corresponds to the Eu$_{\rm Sr}^+$ $\rightarrow$ Eu$_{\rm Sr}^0$ + $h^+$ process [$E_{\rm abs} =$ 3.80 or 3.81 eV; Fig.~\ref{fig;fe;cc}(b)] in our work. On the basis of our results, however, that excitation band should {\it instead} be assigned to the Eu$_{\rm Sr}^0$ $\rightarrow$ Eu$_{\rm Sr}^+$ + $e^-$ process [$E_{\rm abs} =$ 4.86 or 4.71 eV; Fig.~\ref{fig;fe;cc}(a)]. Note that there have been no reports of broad CT emission bands; and only sharp $4f$--$4f$ transitions are observed in the emission spectrum of Eu$^{3+}$ in SrAl$_2$O$_4$ \cite{Botterman2014PRB,Bierwagen2021JL}. This appears to be consistent with our discussion regarding alternative emission processes.

Figure \ref{fig;fe;cc}(c) illustrates similar processes, but now involving the electric-dipole allowed $5d$--$4f$ transitions in the neutral defect configuration Eu$_{\rm Sr}^0$ and with the energies obtained from constrained occupancy HSE calculations. In the absorption process, Eu$_{\rm Sr}^0$ ($4f^7$) absorbs a photon and becomes Eu$_{\rm Sr}^{0,\ast}$ ($4f^65d^1$) with an electron being excited to a higher lying level which is the lowest Eu $5d^1$ state (that is now pushed down from the conduction band due to the occupation of the excited electron). The peak absorption energy related to the $4f^7$ $\rightarrow$ $4f^65d^1$, i.e., the total-energy difference between Eu$_{\rm Sr}^0$ ($4f^7$) and the excited configuration Eu$_{\rm Sr}^{0,\ast}$ ($4f^65d^1$) in the lattice geometry of the former, is 3.91 eV (3.69 eV), with a relaxation energy $d_{\rm FC}^{\rm e}$ = 0.22 eV (0.28 eV), when Eu is incorporated at the Sr1 (Sr2) site. The ZPL, 3.69 eV (3.41 eV) at the Sr1 (Sr2) site, is the total-energy difference between the ground state $4f^7$ [Fig.~\ref{fig;chgdiff}(a)] and the excited state $4f^65d^1$ [Fig.~\ref{fig;chgdiff}(b)]. In the reverse process, the excited electron recombines radiatively with the hole that has been left behind and emits a photon. The peak emission energy related to the $4f^65d^1$ $\rightarrow$ $4f^7$ process, i.e., the total-energy difference between the excited Eu$_{\rm Sr}^{0,\ast}$ ($4f^65d^1$) and the ground state Eu$_{\rm Sr}^0$ ($4f^7$) in the lattice geometry of the former, is 3.42 eV (3.05 eV), with a relaxation energy $d_{\rm FC}^{\rm g}$ = 0.27 eV (0.36 eV), when Eu is at the Sr1 (Sr2) site. The Stokes shift, $\Delta S = d_{\rm FC}^{\rm e} + d_{\rm FC}^{\rm g}$, is calculated to be 0.49 (0.64) eV at the Sr1 (Sr2) site. 

Although there are differences between our calculated energies for the $4f^7$ $\rightleftharpoons$ $4f^65d^1$ processes (summarized in Table \ref{tab;opt}) and those obtained in experiments, the emissions at the Sr1 and Sr2 lattice sites can be identified with the two broad emission bands peaking at 445 nm (2.79 eV, blue) and 520 nm (2.38 eV, green) and the Stokes shifts of roughly 3000 cm$^{-1}$ (0.37 eV) and 4000 cm$^{-1}$ (0.50 eV), respectively, observed in Eu$^{2+}$-doped SrAl$_2$O$_4$ \cite{Poort1995CM,Botterman2014PRB,Bierwagen2016OME}. We find that the difference between the two emission energies is 0.37 eV, just like in experiments (which is larger than that for the band--defect optical transitions discussed earlier, thus indicating the $5d$--$4f$ transitions are more sensitive to the local lattice environments). The calculated peak emission energy and Stokes shift are higher than the reported experimental values by 0.63 eV (0.67 eV) and 0.12 eV (0.14 eV) at the Sr1 (Sr2) site, respectively, which is an almost constant shift among the two lattice sites. The discrepancies with experiments may be ascribed to the electron--hole interaction that is not included in the constrained occupancy approach we employ to describe the excited state of Eu$^{2+}$. 

Note that, using constrained occupancy DFT$+$$U$ calculations, Jia et al.~\cite{Jia2017PRB} reported much lower values for the emission energies. For example, they obtained 2.316 (2.547) eV at the Sr1 (Sr2) site and assigned the experimentally observed green and blue luminescence bands to the Sr1 and Sr2 sites, respectively, which is different from our assignment and that of Ning et al~.\cite{Ning2018JMCC} based on multiconfigurational and constrained occupancy calculations. Our attempts to reproduce the results of Jia et al.~\cite{Jia2017PRB} using a similar computational setup are not successful. Specifically, in our calculations based on DFT$+$$U$ \cite{anisimov1991} with $U$ = 7.5 eV applied on the Eu $4f$ states, a 1$\times$1$\times$2 (56-atom) supercell, and a 3$\times$3$\times$3 $k$-point mesh, we find the emission energy is 3.52 (3.38) eV at the Sr1 (Sr2) site, which shows the same trend as in our HSE-based calculations discussed earlier; see also Table \ref{tab;opt}.

\subsection{Carrier traps for persistent luminescence}\label{sec;results;traps}

Let us now identify charge carrier traps that can play a role in the persistent luminescence of Eu$^{2+}$-doped SrAl$_2$O$_4$. Among the native point defects, we find that Sr$_i$ can act as a trapping center for electrons. Being stable as Sr$_i^{2+}$ in as-prepared SrAl$_2$O$_4$, the defect can capture up to two electrons. The thermodynamic transition levels $(2+/+)$ and $(+/0)$ of Sr$_i$, at 0.44 eV and 0.30 eV below the CBM, respectively, are sufficient close to the CBM, and the positively charged carrier-capturing configurations, i.e., Sr$_i^{2+}$ and Sr$_i^+$, are electrostatically attractive to electrons from the conduction band. 

Note that, in general, the carrier capture cross section increases by orders of magnitude in going from Coulomb repulsive defect centers to neutral centers to attractive centers \cite{Mayer1990}. Also note that the error bar in our calculations of the transitions levels, $\epsilon(q/q')$ and $E_{\rm opt}^{q/q'}$, is about 0.1 eV. A discrepancy of about 0.2 eV should be expected in a comparison between the calculated and the experimental values {\it when} the energy levels are measured from the CBM; here, the additional 0.1 eV is to take into account a possible difference between the calculated and the actual band gaps and/or measurement uncertainties. 

The presence of the Sr$_i$-related trapping centers can explain why the emission observed in Eu$^{2+}$-doped SrAl$_2$O$_4$ is persistent (albeit with a short afterglow time) even without Dy$^{3+}$ co-doping \cite{Abbruscato1971ECS,Matsuzawa1996JES,Lu2007GPC,Bierwagen2020JL}. Notably, our results indicate that oxygen vacancies cannot act as efficient electron traps for room-temperature persistent luminescence (even when they occur with a high concentration, e.g., in samples prepared under reducing conditions) as their defect levels are too deep in the host band gap.  

Finally, with the $(+/0)$ level located at 0.30 (0.38) eV below the CBM at the Sr1 (Sr2) site as reported earlier, Dy$_{\rm Sr}$ can be an efficient electron trap. The electron-capturing configuration, Dy$_{\rm Sr}^+$, is positively charged. Note that, unlike Sr$_i$ where the defect state associated with Sr$_i^+$ and Sr$_i^0$ (i.e., Sr$_i^{2+}$ after capturing one and two electrons, respectively) is derived largely from the host states at the CBM and delocalized over several lattice sites, that associated with Dy$_{\rm Sr}^0$ (i.e., Dy$_{\rm Sr}^+$ after capturing an electron) is highly localized Dy $5d$ states. This indicates that the Dy$_{\rm Sr}$-related traps are much more stable than the Sr$_i$-related ones, which is consistent with the fact that the performance of the afterglow in Eu$^{2+}$-doped SrAl$_2$O$_4$ is significantly improved by Dy co-doping \cite{Katsumata2006JACerS,Nakazawa2006JAP,Bierwagen2020JL}. First-principles calculations of photoionization and carrier capture rates \cite{Razinkovas2021PRB,Dreyer2020PRB} can provide a more quantitative understanding. Also note that the presence of the (Sr$_i$ and Dy$_{\rm Sr}$ related) electron traps is consistent with the fact that the Eu$^{2+}$ $4f^65d^1$ $\rightarrow$ $4f^7$ emission was observed to be quenched via the conduction band \cite{Ueda2012PSSC}.

\section{Conclusions}

We have carried out a comprehensive study of native point defects and rare-earth (co)dopants in SrAl$_2$O$_4$. The major conclusions can be summarized as follows:

1. Eu is mixed valence of Eu$^{2+}$ and Eu$^{3+}$ and energetically most favorable at the Sr sites. The Eu$^{2+}$/Eu$^{3+}$ ratio can be tuned by tuning the synthesis conditions. Similarly, both Dy$^{2+}$ and Dy$^{3+}$ can be stabilized and are energetically most favorable at the Sr sites. Dy$^{2+}$ is, however, always energetically much less favorable than Dy$^{3+}$ and thus would not be realized in synthesis, although it can be photogenerated under irradiation. 

2. Band--defect and interconfigurational $5d$--$4f$ optical transitions involving the Eu$_{\rm Sr}$ defect are investigated using first-principles defect calculations and a constrained occupancy approach, and alternative processes are discussed. On the basis of our results, we assign the broad blue (445 nm) and green (520 nm) emission bands observed in Eu$^{2+}$-doped SrAl$_2$O$_4$ to the Eu$^{2+}$ $4f^65d^1$ $\rightarrow$ $4f^7$ transition at the Sr1 and Sr2 sites, respectively. 

3. Strontium interstitials are found to be efficient electron traps for room-temperature persistent luminescence. When the material is co-doped with Dy, the co-dopant provides an even more stable electron trapping center due to the stabilization of Dy$^{2+}$ which can explain the significantly improved performance of the afterglow in (Eu,Dy)-doped samples. Oxygen vacancies cannot be efficient electron traps, in contrast to what is commonly believed, due to their very deep defect levels.

Our work thus calls for a re-assessment of certain assumptions regarding specific defects previously made in all the mechanisms proposed for the persistent luminescence observed in Eu- and (Eu,Dy)-doped SrAl$_2$O$_4$. It also shows a need to go beyond a constrained occupancy approach in order to obtain more quantitative results for the interconfigurational $5d$--$4f$ optical transitions.  

\begin{acknowledgments}

The author gratefully acknowledges St\'{e}phane Jobic (Nantes Universit\'{e}, CNRS) for helpful discussion. This work used resources of the Center for Computationally Assisted Science and Technology (CCAST) at North Dakota State University, which were made possible in part by NSF MRI Award No.~2019077. 

\end{acknowledgments}

\appendix

\section{Supporting figures and tables}\label{sec;app}

\begin{figure}[h!]
\centering
\includegraphics[width=5.8cm]{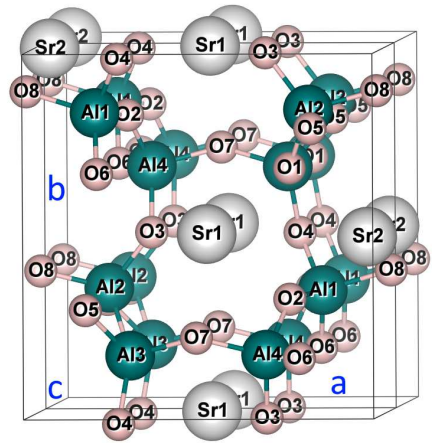}
\caption{Crystal structure of monoclinic SrAl$_2$O$_{4}$ (space group $P2_1$). The unit cell is doubled along the $c$-axis to show the Sr1 and Sr2 channels. Large (gray) spheres are Sr, medium (blue) are Al, and small (red) are O.}
\label{fig;unitcell}
\end{figure}

\begin{table*}[h!]
\caption{Formation enthalpies (calculated at 0 K, in eV per formula unit) of SrAl$_2$O$_4$, Sr--Al--O phases that define the stability region of SrAl$_2$O$_4$, and Eu- and Dy-related phases employed in the determination of the Eu and Dy chemical potentials.}\label{tab;enthalpies}
\begin{center}
\begin{ruledtabular}
\begin{tabular}{rrrrrrrrr}
SrAl$_2$O$_4$ &Sr$_3$Al$_2$O$_6$ &Sr$_4$Al$_{14}$O$_{25}$ & SrO$_2$ & SrAl$_2$ & SrAl$_4$ & Eu$_2$O$_3$ & EuO & Dy$_2$O$_3$ \\
\colrule
$-22.6974$ & $-34.2968$ & $-139.7826$ & $-5.8455$ & $-0.8403$ & $-1.2252$ & $-14.3794$ & $-6.2000$ & $-18.7462$
\end{tabular}
\end{ruledtabular}
\end{center}
\begin{flushleft}
\end{flushleft}
\end{table*}

\begin{figure*}[h!]
\centering
\includegraphics[width=9cm]{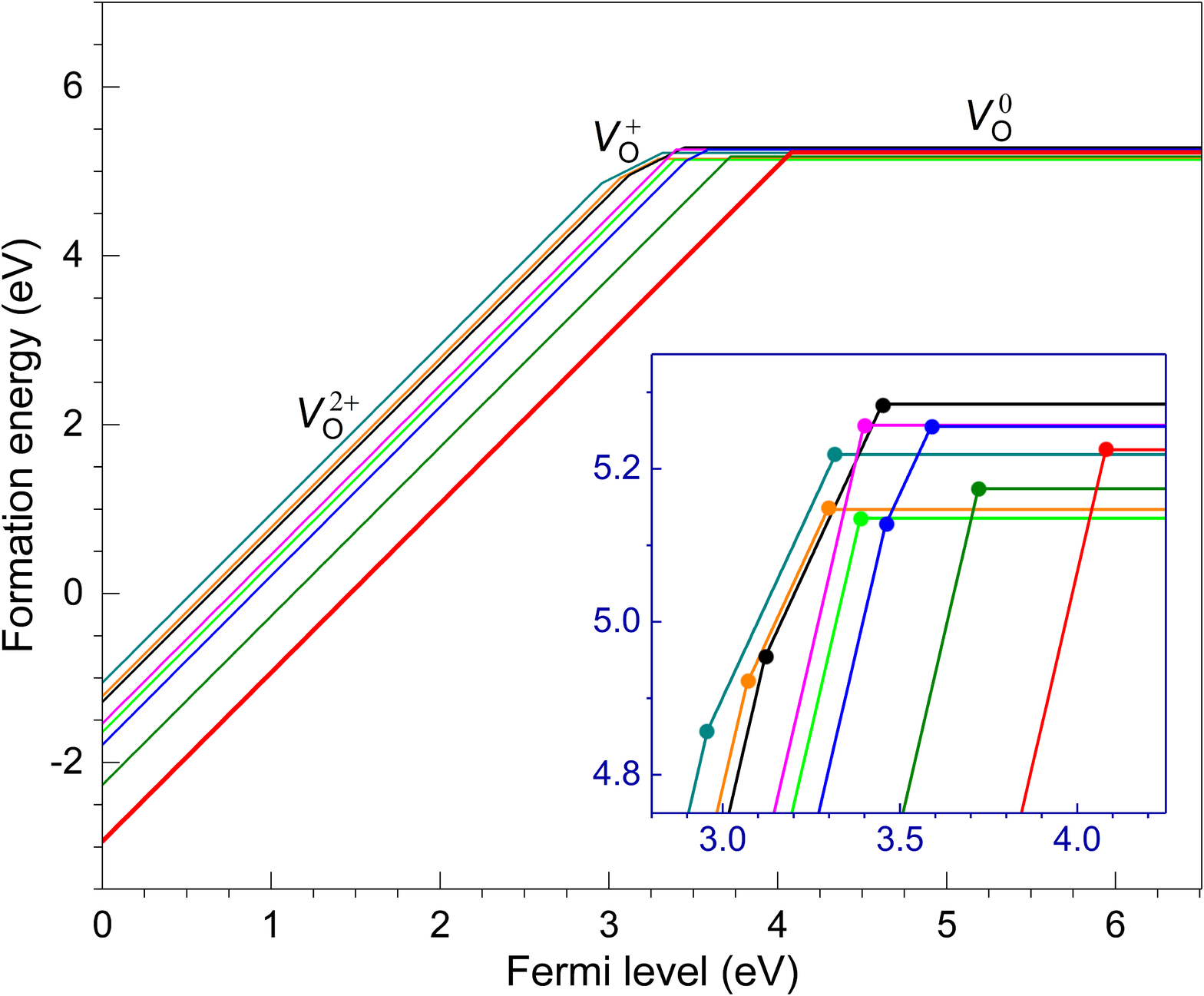}
\caption{Formation energies of oxygen vacancies at eight inequivalent O lattice sites in SrAl$_2$O$_4$ (see Fig.~S1), as a function of the Fermi level from the VBM to the CBM, calculated at point $O1$ in the phase diagram (Fig.~1). For each defect, only segments of the formation energy lines corresponding to the lowest-energy charge states are shown. The kinks connecting two energy segments with different slopes mark the {\it defect levels} [i.e., thermodynamic transition levels $\epsilon(q/q')$]. The defect levels introduced by the vacancies are 2.43--3.55 eV below the CBM. The lowest-energy $V_{\rm O}^{2+}$ configuration occurs at the O8 site.}
\label{fig;vo}
\end{figure*}

\begin{table*}[h!]
\caption{Defect energy levels (in eV, with respect to the VBM) induced by native defects and rare-earth (RE) impurities.}\label{tab;defectlevel}
\begin{center}
\begin{ruledtabular}
\begin{tabular}{llll}
Defect &Lattice site & Stable RE ions & Defect energy levels \\
\colrule
$V_{\rm O}$ & O8 site && $\epsilon(2+/0) = 4.08$ \\
O$_i$ &&& $\epsilon(+/0) = 1.29$, $\epsilon(0/2-) = 3.29$$^a$ \\
$V_{\rm Sr}$ & Sr1 site && $\epsilon(+/0) = 0.95$, $\epsilon(0/-) = 1.21$, $\epsilon(-/2-) = 1.73$ \\
             & Sr2 site && $\epsilon(+/0) = 0.94$, $\epsilon(0/-) = 1.24$, $\epsilon(-/2-) = 1.73$ \\
Sr$_i$ & Sr1 channel && $\epsilon(3+/2+) = 0.61$, $\epsilon(2+/+) = 6.00$, $\epsilon(+/0) = 6.17$ \\
       & Sr2 channel && $\epsilon(3+/2+) = 0.65$, $\epsilon(2+/+) = 6.08$, $\epsilon(+/0) = 6.21$ \\
Al$_{\rm Sr}$ & Sr2 site && $\epsilon(2+/+) = 0.24$, $\epsilon(+/-) = 5.10$$^b$ \\
Sr$_{\rm Al}$ & Al2 site && $\epsilon(+/0) = 1.37$, $\epsilon(0/-) = 1.80$ \\
Eu$_{\rm Sr}$ & Sr1 site &Eu$^{3+}$, Eu$^{2+}$ & $\epsilon(+/0) = 2.69$ \\
              & Sr2 site &Eu$^{3+}$, Eu$^{2+}$ & $\epsilon(+/0) = 2.79$ \\
Eu$_{\rm Al}$ & Al2 site &Eu$^{3+}$, Eu$^{2+}$ & $\epsilon(2+/+) = 0.75$, $\epsilon(+/0) = 0.86$, $\epsilon(+/0) = 4.87$ \\							
Dy$_{\rm Sr}$ & Sr1 site &Dy$^{3+}$, Dy$^{2+}$ & $\epsilon(+/0) = 6.21$ \\
              & Sr2 site &Dy$^{3+}$, Dy$^{2+}$ & $\epsilon(+/0) = 6.13$ \\
Dy$_{\rm Al}$ & Al2 site &Dy$^{4+}$, Dy$^{3+}$ & $\epsilon(2+/+) = 0.60$, $\epsilon(+/0) = 1.07$ \\		
\end{tabular}
\end{ruledtabular}
\end{center}
\begin{flushleft}
$^a$The $\epsilon(0/-)$ and $\epsilon(-/2-)$ levels are very close to the $\epsilon(0/2-)$ level, at 3.28 and 3.30 eV, respectively. 
 
$^b$The $\epsilon(+/0)$ and $\epsilon(0/-)$ levels are very close to the $\epsilon(+/-)$ level, at 5.10 and 5.09 eV, respectively.  
\end{flushleft}
\end{table*}

\begin{figure*}[h!]
\vspace{0.2cm}
\includegraphics*[width=0.75\linewidth]{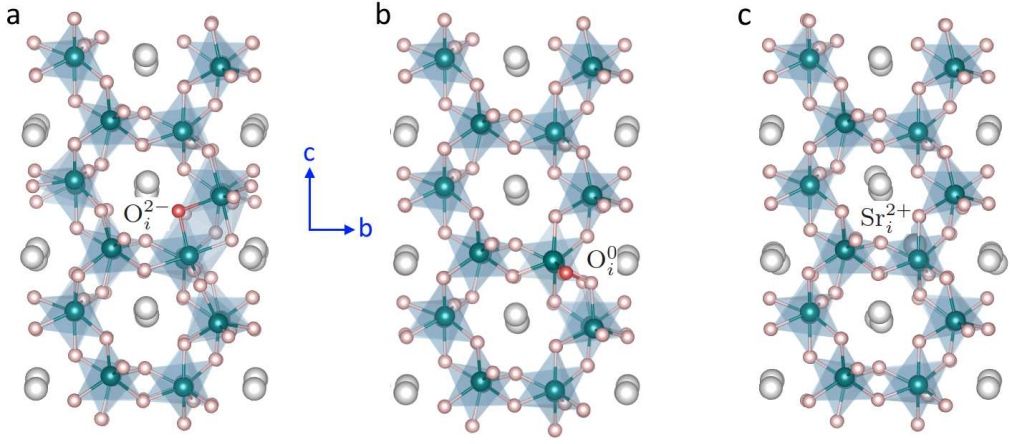}
\caption{A different view of (a) O$_i^{2-}$, (b) O$_i^0$, and (c) Sr$_i^{2+}$ (in the Sr2 channel; between two AlO$_4$ units when viewed along the $a$-axis) defect configurations in SrAl$_2$O$_4$. Large (gray) spheres are Sr, medium (blue) are Al, and small (red) are O.} 
\label{fig;native;struct;extra} 
\end{figure*}

\begin{figure*}[h!]
\centering
\includegraphics[width=16.0cm]{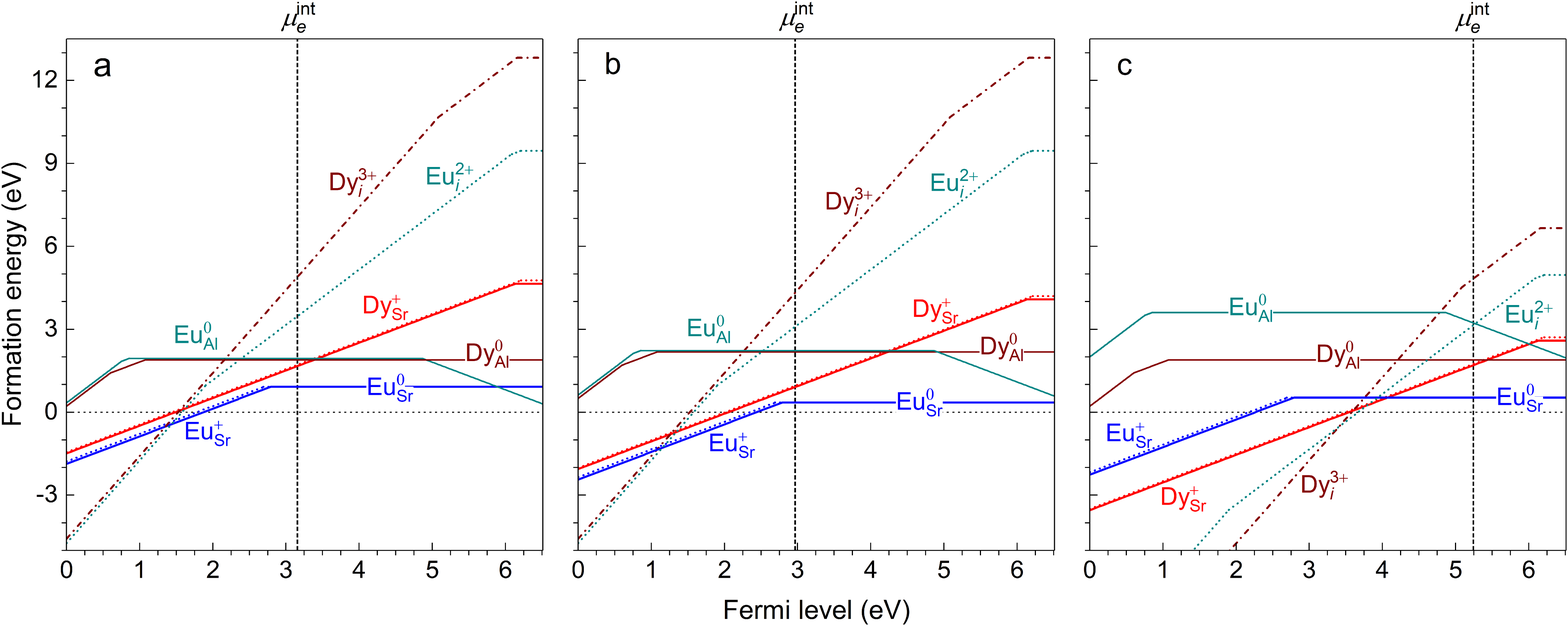}
\caption{Formation energies of Eu- and Dy-related defects in SrAl$_2$O$_4$, calculated at points (a) $O1$, (b) $O2$, and (c) $R$ in the phase diagram (Fig.~1). The results for the Eu and Dy interstitials are also included. $\mu_e^{\rm int}$ is the Fermi-level position determined by native point defects; see the main text. The kinks connecting two segments with different slopes mark the {\it defect levels}.}
\label{fig;re;all}
\end{figure*}

\begin{figure*}[h!]
\vspace{0.2cm}
\includegraphics*[width=0.75\linewidth]{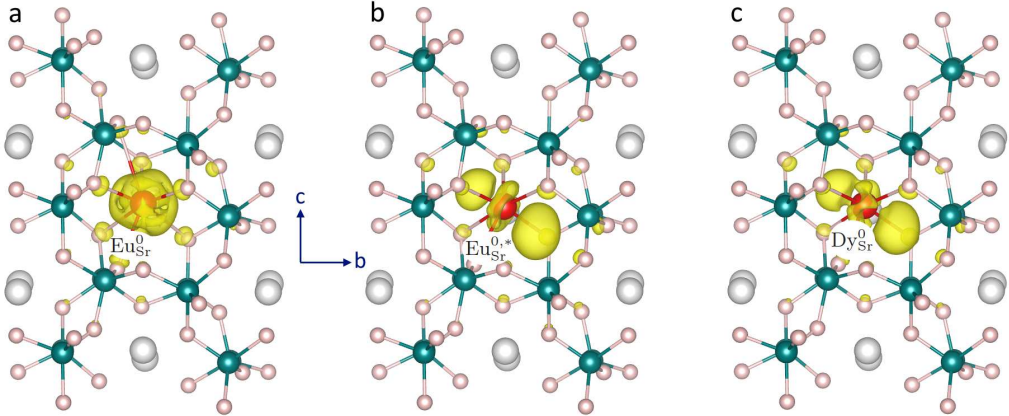}
\caption{Charge densities showing the localized electron residing at the highest occupied state of (a) Eu$_{\rm Sr}^0$ ($4f^7$), (b) Eu$_{\rm Sr}^{0,\ast}$ ($4f^65d^1$), and (c) Dy$_{\rm Sr}^0$ ($4f^95d^1$); all the three defect configurations are at the Sr2 site. The isovalue for the isosurface is set to 0.03 $e$/{\AA}$^3$. Larger (red) spheres are Eu/Dy, large (gray) are Sr, medium (blue) are Al, and small (red) are O.} 
\label{fig;chgdiff} 
\end{figure*}

\begin{table*}[h!]
\caption{Stable charge states of Eu-related defect complexes, their constituent defects, and binding energies ($E_{\rm b}$).}\label{tab;complex}
\begin{center}
\begin{ruledtabular}
\begin{tabular}{llrllr}
Complex & Constituents & $E_{\rm b}$ (eV) & Complex & Constituents & $E_{\rm b}$ (eV)\\
\cmidrule(lr){1-3} \cmidrule(lr){4-6}
(Eu$_{\rm Sr}$-Dy$_{\rm Sr}$)$^{2+}$ & Eu$_{\rm Sr}^+$ + Dy$_{\rm Sr}^+$ & $-$0.41 & (Eu$_{\rm Sr}$-Sr$_i$)$^{3+}$ & Eu$_{\rm Sr}^+$ + Sr$_i^{2+}$ & $-$0.69 \\
(Eu$_{\rm Sr}$-Dy$_{\rm Sr}$)$^{+}$ & Eu$_{\rm Sr}^0$ + Dy$_{\rm Sr}^+$ & $-$0.02 & (Eu$_{\rm Sr}$-Sr$_i$)$^{2+}$ & Eu$_{\rm Sr}^0$ + Sr$_i^{2+}$ & 0.01 \\
(Eu$_{\rm Sr}$-Dy$_{\rm Sr}$)$^{0}$ & Eu$_{\rm Sr}^0$ + Dy$_{\rm Sr}^0$ & 0.01 & (Eu$_{\rm Sr}$-Sr$_i$)$^{+}$ & Eu$_{\rm Sr}^0$ + Sr$_i^{+}$ & 0.05  \\
(Eu$_{\rm Sr}$-$V_{\rm O}$)$^{3+}$ & Eu$_{\rm Sr}^+$ + $V_{\rm O}^{2+}$ & $-$0.73 & (Eu$_{\rm Sr}$-Sr$_i$)$^{0}$ & Eu$_{\rm Sr}^0$ + Sr$_i^{0}$ & 0.05 \\
(Eu$_{\rm Sr}$-$V_{\rm O}$)$^{2+}$ & Eu$_{\rm Sr}^0$ + $V_{\rm O}^{2+}$ & 0.07 & (Eu$_{\rm Sr}$-$V_{\rm Sr}$)$^{+}$ & Eu$_{\rm Sr}^+$ + $V_{\rm Sr}^{0}$ & 0.31\\
(Eu$_{\rm Sr}$-$V_{\rm O}$)$^{0}$ & Eu$_{\rm Sr}^0$ + $V_{\rm O}^{0}$ & 0.01 & (Eu$_{\rm Sr}$-$V_{\rm Sr}$)$^{0}$ & Eu$_{\rm Sr}^+$ + $V_{\rm Sr}^{-}$ & 0.61 \\
(Eu$_{\rm Sr}$-O$_i$)$^{+}$ & Eu$_{\rm Sr}^+$ + O$_i^{0}$ & $-$0.06 & (Eu$_{\rm Sr}$-$V_{\rm Sr}$)$^{-}$ & Eu$_{\rm Sr}^+$ + $V_{\rm Sr}^{2-}$ & 1.15 \\
(Eu$_{\rm Sr}$-O$_i$)$^{-}$ & Eu$_{\rm Sr}^+$ + O$_i^{2-}$ & 1.09 & (Eu$_{\rm Sr}$-$V_{\rm Sr}$)$^{2-}$ & Eu$_{\rm Sr}^0$ + $V_{\rm Sr}^{2-}$ & 0.07 \\
(Eu$_{\rm Sr}$-O$_i$)$^{2-}$ & Eu$_{\rm Sr}^0$ + O$_i^{2-}$ & 0.00 \\
     
\end{tabular}
\end{ruledtabular}
\end{center}
\begin{flushleft}
\end{flushleft}
\end{table*}

\begin{figure*}[h!]
\centering
\includegraphics[width=9.0cm]{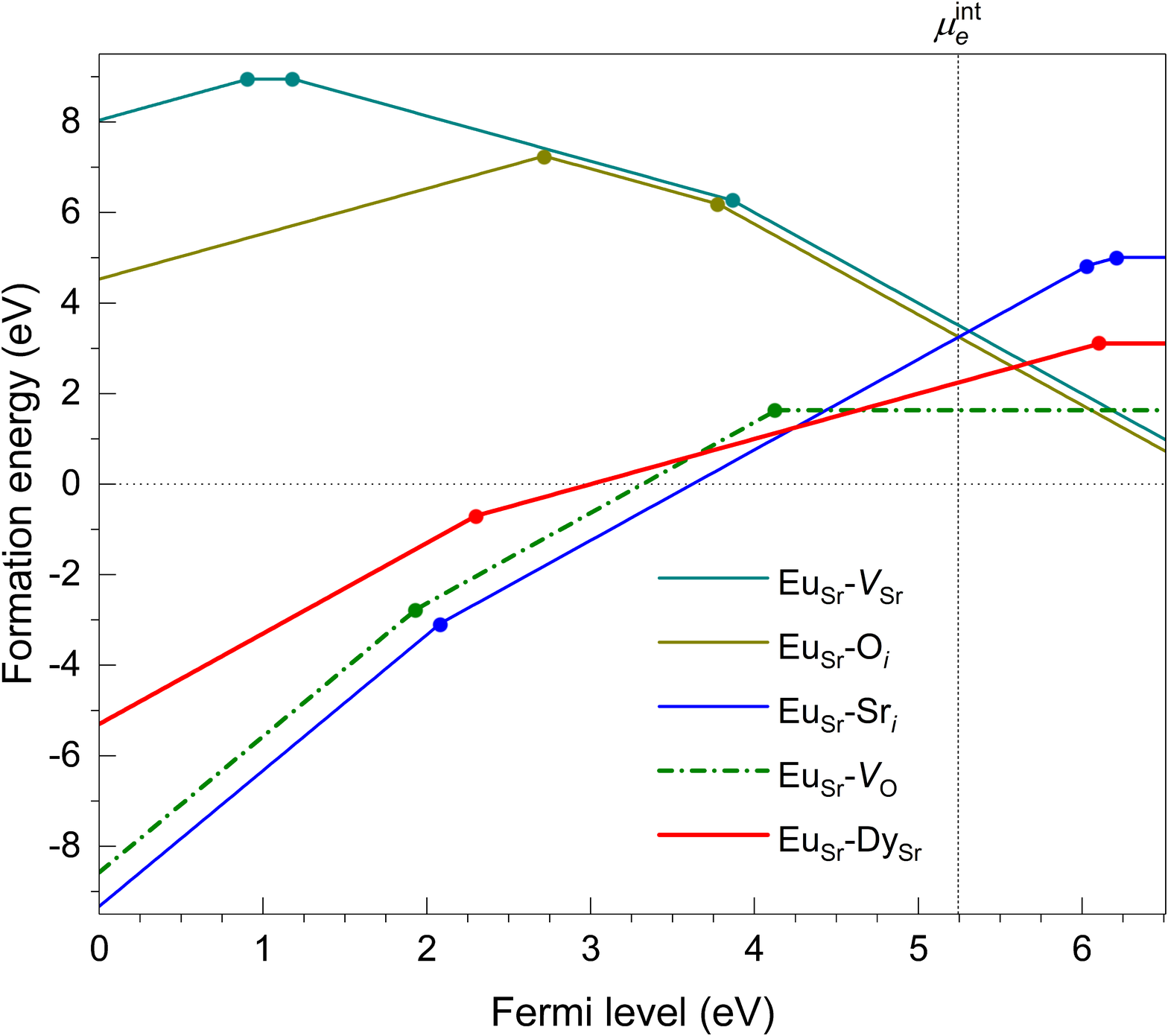}
\caption{Formation energies of possible Eu-related defect complexes in SrAl$_2$O$_4$, calculated at point $R$ in the phase diagram (Fig.~1). The two constituent defects in a complex are nearest neighbors to each other. $\mu_e^{\rm int}$ is the Fermi-level position determined by native point defects. The solid dots connecting two energy segments with different slopes mark the {\it defect levels}.}
\label{fig;complex}
\end{figure*}

\begin{table*}[h!]
\caption{Peak absorption energy ($E_{\rm abs}$), peak emission energy ($E_{\rm m}$), Franck-Condon shifts ($d_{\rm FC}^{\rm e,g}$), Stokes shift ($\Delta S$), and thermal energy ($E_{\rm therm}$) associated with the Eu-related band--defect and $5d$--$4f$ optical transitions; all in eV.}\label{tab;opt}
\begin{center}
\begin{ruledtabular}
\begin{tabular}{ccccccccccccc}
Lattice site & $E_{\rm abs}$ & $d_{\rm FC}^{\rm e}$ & $E_{\rm em}$ & $d_{\rm FC}^{\rm g}$  & $\Delta S$ & $E_{\rm therm}$ &$E_{\rm abs}$ & $d_{\rm FC}^{\rm e}$ & $E_{\rm em}$ & $d_{\rm FC}^{\rm g}$  & $\Delta S$ & $E_{\rm therm}$ \\
\cmidrule(lr){1-1}\cmidrule(lr){2-7} \cmidrule(lr){8-13}
& \multicolumn{6}{c}{HSE: Eu$_{\rm Sr}^0$ $\rightleftharpoons$ Eu$_{\rm Sr}^+$ $+$ $e^-$} & \multicolumn{6}{c}{HSE: Eu$_{\rm Sr}^0$ ($4f^7$) $\rightleftharpoons$ Eu$_{\rm Sr}^{0,\ast}$ ($4f^65d^1$)} \\
\cmidrule(lr){2-7} \cmidrule(lr){8-13}
Sr1 & 4.86 & 1.04 & 2.71 & 1.11 & 2.15 & 3.82 & 3.91 & 0.22 & 3.42 & 0.27 & 0.49 & 3.69  \\
Sr2 & 4.71 & 0.98 & 2.70 & 1.02 & 2.00 & 3.72 & 3.69 & 0.28 & 3.05 & 0.36 & 0.64 & 3.41 \\
\cmidrule(lr){2-7} \cmidrule(lr){8-13}
& \multicolumn{6}{c}{HSE: Eu$_{\rm Sr}^+$ $\rightleftharpoons$ Eu$_{\rm Sr}^0$ $+$ $h^+$}& \multicolumn{6}{c}{DFT$+$$U$:$^a$ Eu$_{\rm Sr}^0$ ($4f^7$) $\rightleftharpoons$ Eu$_{\rm Sr}^{0,\ast}$ ($4f^65d^1$)} \\
\cmidrule(lr){2-7} \cmidrule(lr){8-13}
 Sr1 & 3.80 & 1.11 & 1.65 & 1.04 & 2.15 & 2.69 & 4.23 & 0.32 & 3.52 & 0.39 & 0.61 &3.91 \\
 Sr2 & 3.81 & 1.02 & 1.81 & 0.98 & 2.00 & 2.79 & 4.00 & 0.29 & 3.38 & 0.33 & 0.72 &3.71   
\end{tabular}
\end{ruledtabular}
\end{center}
\begin{flushleft}
$^a$DFT+$U$ calculations with a computational setup similar to that in Jia et al., Phys.~Rev.~B {\bf 96}, 125132 (2017); see text.
\end{flushleft}
\end{table*}

%

\end{document}